\documentclass[a4paper,10pt]{amsart}

\usepackage{amsthm}
\numberwithin{equation}{section}
\newtheorem{theorem}{Theorem}[section]

\theoremstyle{definition}
\newtheorem{remark}[theorem]{Remark}
\newtheorem{definition}[theorem]{Definition}

\usepackage{amsmath,amsfonts,amsbsy}

\usepackage{color}
\usepackage[colorlinks=true,linkcolor=blue,urlcolor=blue,citecolor=blue]{hyperref}

\usepackage{dsfont}

\usepackage{graphicx,subcaption}
\usepackage[font=footnotesize]{caption}
\graphicspath{ {Plots/} }

\usepackage[latin1]{inputenc}  

\newcommand{\eps}{\varepsilon}
\newcommand{\Id}{\mathds{1}}  
\newcommand{\id}{\mathbb{I}}   
\newcommand{\B}{\mathbb{B}}

\newcommand{\C}{\mathbb{C}}
\newcommand{\R}{\mathbb{R}}
\newcommand{\Z}{\mathbb{Z}}
\newcommand{\N}{\mathbb{N}}
\newcommand{\T}{\mathbb{T}}

\newcommand{\F}{\mathcal{F}}
\newcommand{\bra}[1]{\left\langle #1 \right|}
\newcommand{\ket}[1]{\left| #1 \right\rangle}
\newcommand{\eu}{\mathrm{e}}
\newcommand{\iu}{\mathrm{i}}
\newcommand{\di}{\mathrm{d}}

\newcommand{\sub}[1]{_{\mathrm{#1}}}

\newcommand{\crucial}[1]{{\it \textbf{#1}}}

\newcommand{\inner}[2]{\left\langle #1, #2 \right\rangle}
\newcommand{\norm}[1]{\left\| #1 \right\|}

\newcommand{\half}{\mbox{\footnotesize $\frac{1}{2}$}}  
\newcommand{\iso}{\cong}   

\newcommand{\ve}[1]{\mathbf{#1}}  
\newcommand{\V}[1]{\mathbf{#1}}   

\newcommand{\cry}{\mathcal{C}}     

\newcommand{\la}{\lambda}

\newcommand{\ph}{\varphi}

\def\({\left(}
\def\){\right)}

\DeclareMathOperator{\Tr}{Tr}
\DeclareMathOperator{\Ran}{Ran}
\DeclareMathOperator{\dist}{dist}

\newcommand{\ie}{{\sl i.\,e.\ }}
\newcommand{\eg}{{\sl e.\,g.\ }}

\newcommand{\set}[1]{ \left\{  #1 \right\}}

\newcommand{\Lattice}{\mathfrak{D}}
\newcommand{\x}{\mathbf{x}}
\newcommand{\y}{\mathbf{y}}
\newcommand{\z}{\mathbf{z}}
\newcommand{\w}{\psi}
\newcommand{\Loc}{G}

\usepackage{enumerate}
\usepackage{enumitem}
\renewcommand\labelenumi{(\roman{enumi})}
\renewcommand\theenumi\labelenumi

\let\oldfootnote\footnote
\renewcommand{\footnote}[1]{\oldfootnote{\  #1}}

\title[The Haldane model and its localization dichotomy]{The Haldane model\\[1mm] and its localization dichotomy}

\author[G.~Marcelli]{Giovanna Marcelli}
\address[G.~Marcelli]{Fachbereich Mathematik, Eberhard Karls Universit\"{a}t T\"{u}bingen. Auf der Morgenstelle 10, 72076 T\"{u}bingen (DE).}
\email{\href{mailto:giovanna.marcelli@uni-tuebingen.de}{\texttt{giovanna.marcelli@uni-tuebingen.de}}.}

\author[D.~Monaco]{Domenico Monaco}
\address[D.~Monaco]{Dipartimento di Matematica e Fisica, Universit\`{a} degli Studi di Roma Tre. Largo San Leonardo Murialdo 1, 00146 Rome (IT).}
\email{\href{mailto:dmonaco@mat.uniroma3.it}{\texttt{dmonaco@mat.uniroma3.it}}.}

\author[M.~Moscolari]{Massimo Moscolari}
\address[M.~Moscolari]{Dipartimento di Matematica, ``La Sapienza'' Universit\`{a} di Roma. \mbox{Piazzale} Aldo Moro 2, 00185 Rome (IT) \newline \textsl{and} \newline 
Department of Mathematical Sciences, Aalborg University. Skjernvej 4A, 9220 \mbox{Aalborg} (DK)}
\email{\href{mailto:moscolari@mat.uniroma1.it}{\texttt{moscolari@mat.uniroma1.it}}.}

\author[G.~Panati]{Gianluca Panati}
\address[G.~Panati]{Dipartimento di Matematica, ``La Sapienza'' Universit\`{a} di Roma. \mbox{Piazzale} Aldo Moro 2, 00185 Rome (IT).}
\email{\href{mailto:panati@mat.uniroma1.it}{\texttt{panati@mat.uniroma1.it}}.}

\begin{document}
\renewcommand{\subjclassname}{%
\textup{2010} Mathematics Subject Classification}

\keywords{Periodic Schr\"{o}dinger operators, Chern insulators, Haldane model, 
Quantum Anomalous Hall Effect, Bloch frames, Wannier functions}
\subjclass[2010]{81Q70,  	
81V70,  	
47A56, 
47A10.
}

\date{ September 6, 2019. Extended version of the paper published in Rend. Mat. Appl. {\bf 39}, 307-327 (2018). In comparison with the published version, we added some details and the whole Chapter 5.}

\maketitle

\begin{center}
Dedicated to Gianfausto, on the occasion of his $85^{\rm th}$ birthday.
\end{center}

\begin{abstract} Gapped periodic quantum systems exhibit an interesting Localization Dichotomy, 
which emerges when one looks at the localization of the optimally localized Wannier functions associated 
to the Bloch bands below the gap. As recently proved, either these Wannier functions are exponentially localized, as it happens whenever  the Hamiltonian operator is time-reversal symmetric, or they are delocalized in the sense that the expectation 
value of  $|\ve{x}|^2$ diverges. Intermediate regimes are forbidden. 

Following the lesson of our {\it Maestro}, to whom this contribution is gratefully dedicated, we find useful to explain this subtle mathematical phenomenon in the simplest possible model, namely the discrete model proposed by Haldane \cite{Haldane88}. 
We include a pedagogical introduction to the model and we explain its Localization Dichotomy by explicit analytical arguments. We then introduce the reader to the more general, model-independent version of the dichotomy proved in  
\cite{MonacoPanatiPisanteTeufel2018}, and finally we announce further generalizations to non-periodic models. 
\end{abstract}



\section{Introduction}
\label{Sec:Introduction}

Gianfausto Dell'Antonio has been always transmitting to younger collaborators the attitude 
to understand -- and explain --  a mathematical phenomenon in the simplest possible model which 
still captures its essential features. Remembering his recommendation, we devote this contribution 
to explain a recent, model-independent result -- namely the \emph{Localization Dicothomy}  for 
gapped periodic quantum systems, proved in \cite{MonacoPanatiPisanteTeufel2018} -- 
by illustrating its essential features in a simple, but yet physically relevant, discrete model.  

We consider the model proposed by Haldane in \cite{Haldane88}, which has become one of the paradigmatic models to describe Chern insulators, a subclass of topological insulators \cite{Ando, HasanKane, FruchartCarpentier2013}. 
Haldane argued that the essential ingredient in the Quantum Hall Effect (QHE)  is the breaking of time-reversal symmetry, an effect that can be obtained either by an external magnetic field (as in a QHE setup)  or, alternatively, by some mechanism internal to the sample, 
as \eg the presence of strong magnetic dipole moments of the ionic cores.   
In Haldane's words \cite{Haldane88}:  
\begin{quote}
``{\it
While the particular model presented here is unlikely to be directly physically realizable, it indicates that,
at least in principle, {the QHE can be placed in the wider context of phenomena associated with 
broken time-reversal invariance, and does not necessarily require external magnetic fields}, but could occur as a consequence 
of magnetic ordering in a quasi-two-dimensional system.}'' 
\end{quote}

\noindent Remarkably, the first sentence turned out to be too pessimistic:  after three decades, Chern insulators predicted in \cite{Haldane88} have been experimentally synthesized as crystalline solids \cite{Experiment,Bestwick et al 2015, Chang et al 2015} and the Haldane model can also be physically simulated by Bose-Einstein condensates in suitably arranged optical lattices.  

In this paper, we first provide a pedagogical introduction to the Haldane model, which is here presented in the first-quantization formalism, as opposed to most of the physics literature, which uses instead a second-quantization language. 
In Section~\ref{Sec:Singularities}, we recall the definition of {\it Bloch functions} and of the {\it Chern number}  associated to an isolated Bloch band, and 
we exhibit, in the Haldane model, a Bloch function producing a non-zero Chern number and having a singular derivative: more precisely, its $H^1$-norm diverges. This quantitative relation between non-trivial topology and the allowed singularities of Bloch functions in the Haldane model was investigated numerically in \cite{ThonhauserVanderbilt}. This situation exemplifies a recent model-independent mathematical result \cite{MonacoPanatiPisanteTeufel2018}, which shows that a non-zero Chern number indeed forces a divergence of the $H^1$-norm of the corresponding Bloch functions in any Bloch gauge. 
We explain in Section~\ref{Sec:Dichotomy}  how the latter divergence reflects into the delocalization of the corresponding {\it Wannier functions}, and we illustrate to the reader the more general {\it Localization Dichotomy} mentioned above. 
A natural question is whether the previous result -- whose formulation heavily relies on periodicity --  can be recast in the broader context of non-periodic models. Some preliminary results in this direction, still unpublished \cite{MarcelliMoscolariPanati}, are announced in Section~\ref{Sec:GeneralizedDichotomy}.

We hope that the introductory style of this contribution will be useful to fill the linguistic gap between mathematics and physics, as they represent a unity in the scientific vision of the person to whom the paper is dedicated.   


\bigskip

\noindent \textbf{Dedication.} The senior author of this paper moved his first steps into the scientific world 
under the precious guidance of Gianfausto Dell'Antonio. From his example, 
as a scientist and as a human being, he learned not only how to do mathematics, but how to be a Mathematical Physicist. 
We all -- authors of different generations -- consider Gianfausto as our {\it Maestro}, and we gratefully acknowledge the unvaluable contribution he gave to the development of Quantum Mathematics in Italy over more than half a century.      
 
\bigskip

\noindent \textbf{Acknowledgements.}  
We are grateful to Cl\'ement Tauber for many useful discussions on the related Kane-Mele model, 
and for his precious help with some Figures. 


\section{The Haldane model and its symmetries}
\label{Sec:Haldane}

The tight-binding model proposed by Haldane \cite{Haldane88}  has become a paradigm in solid-state physics, as it is 
presumably the simplest physically-reasonable model which is invariant by lattice-translations (a unitary ${\Z^2}$-symmetry) and simultaneously breaks, for some values of the parameters $(\phi, M)$ labeling the model, time-reversal symmetry (an antiunitary $\Z_2$-symmetry).  In view of that, it has become one of the most popular models to study materials in the Altland-Zirnbauer symmetry class A, which includes Quantum Hall systems and Chern insulators \cite{Ando, HasanKane, FruchartCarpentier2013}.  
The Haldane model is usually presented by using a second-quantization formalism \cite{FruchartCarpentier2013, Santoro_LN,GiulianiMastropietroPorta2017}, which makes it difficult to readers unfamiliar with the latter to appreciate the simplicity and elegance 
of the essential ideas. Since second quantization is not needed at all to describe non-interacting electrons, 
we review in this Section the essential features of the Haldane model, in a pedagogical style, by using the usual language of discrete Schr\"odinger  operators (\ie a first-quantization formalism).   

\medskip

\subsection{The honeycomb structure} 

\begin{figure}[htb]
\centering
\includegraphics{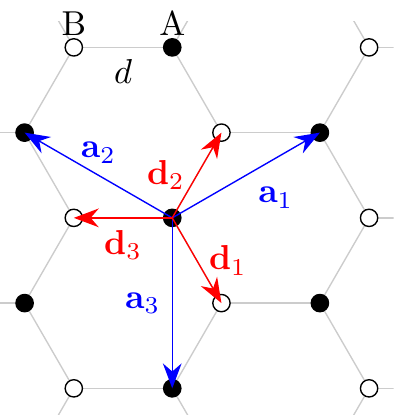}
\caption{\footnotesize The honeycomb structure, with the displacement vectors $\set{\ve d_1, \ve d_2,\ve d_3}$  and the periodicity vectors $\set{\ve a_1, \ve a_2,\ve a_3}$ (color online).
\label{fig:honeycomb}}
\end{figure}

The Haldane model describes independent electrons on a honeycomb structure
\footnote{The physics literature usually refers to the latter as a ``honeycomb lattice''. 
We prefer to avoid here this ambiguous use of the word ``lattice'', since this word has a 
precise meaning in mathematics: a lattice is a discrete subgroup of  $(\R^d, +)$ with maximal rank.
The ambiguity does not arise when speaking about the \emph{Bravais lattice}, which is a lattice for both 
physicists and mathematicians.}\  
 $\mathcal{C}\subset \R^2$, illustrated in Figure \ref{fig:honeycomb}. The structure is characterized by the \emph{displacement vectors} 
$$
\V d_1 = d\begin{pmatrix} \frac{1}{2} & -\frac{\sqrt{3}}{2} \end{pmatrix}, \qquad \V d_2 = d\begin{pmatrix} \frac{1}{2} &\frac{\sqrt{3}}{2} \end{pmatrix}, \qquad \V d_3 = d\begin{pmatrix} -1 & 0 \end{pmatrix} = -\V d_1-\V d_2,
$$
where $d$ is the {smallest} distance between two points of $\mathcal{C}$. 
The periodicity of the structure is expressed by the \emph{periodicity vectors} 
\begin{equation}
\V a_1  = \V d_2 - \V d_3, \qquad \V a_2 = \V d_3-\V d_1, \qquad \V a_3 = \V d_1-\V d_2 = -\V a_1 - \V a_2.
\end{equation}

\begin{figure}[htb]
\centering
\includegraphics[width=\textwidth]{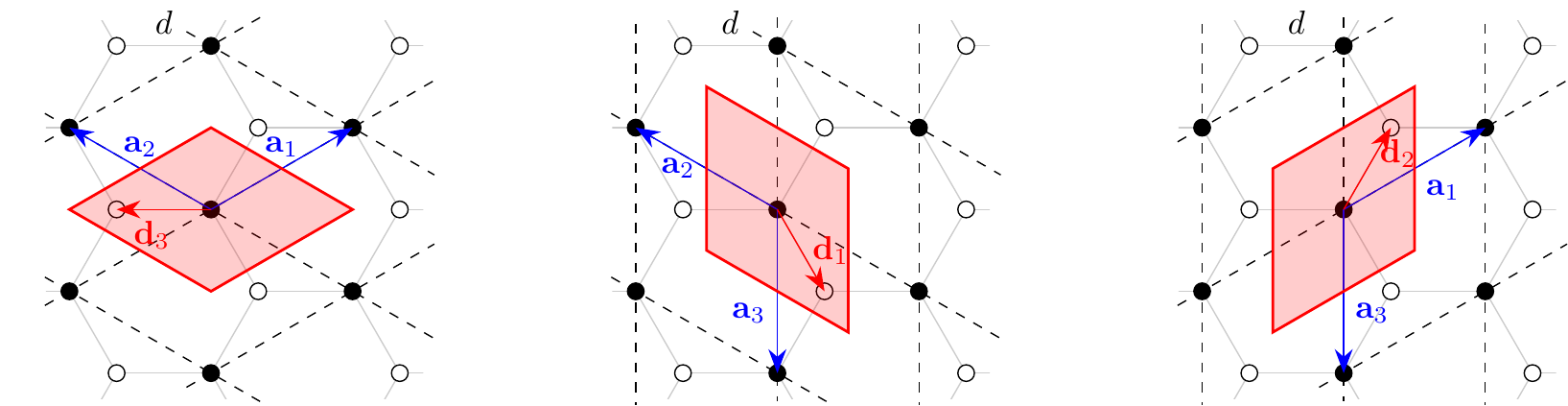}
\caption{\footnotesize Three possible dimerizations of the honeycomb structure, corresponding to three different periodicity cells
(color online). \label{fig:unitcells}}
\end{figure}

The vectors $\V a_i$ generate a Bravais lattice $\Gamma := \mathrm{Span}_\mathbb{Z}\{ \V a_1, \V a_2, \V a_3\} \cong \mathbb Z^2$ where one $\V a_i$ is redundant as it is an integer linear combination of the other two.  
Any point $\x \in \cry$ can be written by using a Bravais lattice vector and one of the $\V d_i$ vectors. It is then sufficient to pick two $\V a_i$-vectors and one $\V d_i$-vector to generate  the whole crystal. This choice,  which is often called a \emph{dimerization} of $\mathcal C$, is not unique, as illustrated in Figure \ref{fig:unitcells}. The above procedure is equivalent to the choice of a  periodicity cell that contains two non-equivalent sites $A$ and $B$ (black and white dots in Figure \ref{fig:honeycomb}, respectively), and is a fundamental cell {w.r.t.\ the action of $\Gamma$}. Hence, each choice of a periodicity cell provides an identification 
$\cry \iso \Gamma \times \set{\ve 0, \ve \nu}$, where $\nu$ is one of the displacement vectors, 
yielding an isomorphism
\footnote{From an abstract viewpoint, we are just using the fact that the $L^2$-functor, from measure spaces to Hilbert spaces, preserves the product structure, mapping the cartesian product into the tensor product.  
} 
$\ell^2(\mathcal C) \cong \ell^2(\Gamma) \otimes \C^2 \iso  \ell^2(\Gamma, \C^2)$. We will often use this ``dimerization
isomorphism'' and the following typographic convention: 
\begin{itemize}
	\item a small letter for a function $\psi \in \ell^2(\cry)$,  with complex values $\psi_\x$ for $\x \in \cry$; 
	\item capital letter for  a function $\Psi \in  \ell^2(\Gamma, \C^2)$;  we make use of a pseudo-spin notation, namely  
                    $$
                    \Psi_{\gamma} = 
                    \left(
                    \begin{array}{c}
                    \psi_{\gamma, A}   \\
                    \psi_{\gamma, B}   
                    \end{array}
                    \right)
                     \qquad \text{ for } \gamma \in \Gamma,
                    $$
         where the labels $A$ and $B$ refer respectively to {the sublattices $\Gamma_A$ and $\Gamma_B$}, 
         so that $\cry = \Gamma_A \cup \Gamma_B$.  
         For example, in the three dimerizations appearing in Figure~2, one has {$\Gamma_A = \Gamma$ and $\Gamma_B = 
         \Gamma + \ve \nu$}, where $\ve \nu \in \set{\ve d_1, \ve d_2,\ve d_3}$ depends on the chosen dimerization. 
\end{itemize}

{Finally, notice that the honeycomb structure has an interesting {\it \textbf{inversion symmetry}}, namely a reflection w.r.t.\ a specific line, which exchanges the role of the sublattices $\Gamma_A$ and $\Gamma_B$. Thus, it yields a $\Z_2$-symmetry which can be easily broken by adding an on-site $\Gamma$-periodic potential which distinguishes between $\Gamma_A$ and $\Gamma_B$.} The latter procedure corresponds to a variation of the parameter $M$ in the Haldane Hamiltonian, to be introduced shortly, and to the transition from graphene to boron-nitride sheets in physical reality.

\goodbreak

\subsection{The  Hamiltonian}
The Haldane model is defined, in a first quantization formalism, through a Hamiltonian operator acting on
$\ell^2(\mathcal C) \iso  \ell^2(\Gamma, \C^2)$, and depending on two real parameters $(\phi, M)$, with 
${\phi \in (-\pi, \pi]}$ representing a magnetic flux and $M \in \R$ corresponding to an on-site energy which distinguishes among the two sublattices {$\Gamma_A$ and $\Gamma_B$}.     

\noindent  The translation operator  $T_{\V u}$, corresponding to a translation by $\ve u \in \R^2$, is defined by
\begin{equation}\label{Translations} 
(T_{\V u} \psi)_{\V x} = \begin{cases}
\psi_{\V x -\V u}             & \text{ if }  {\V x-\V u} \in \mathcal{C}\\
0                                    & \text{ otherwise}
\end{cases} \qquad \text{for all } \psi \in \ell^2(\mathcal{C}). 
\end{equation}
Moreover, we denote by $\chi_A$ (resp.\ $\chi_B$) the charachteristic function of the sublattice {$\Gamma_A$ (resp.\ $\Gamma_B$)}.

Equipped with this notation, one defines the Haldane operator  $H \equiv H_{(\phi, M)}$ acting in 
$\ell^2(\cry)$  
(\ie without reference to a specific dimerization) as a sum of three terms
\begin{equation} \label{Haldane}
{H =  H\sub{NN} + H\sub{NNN} + V .}
\end{equation}
{The nearest neighbor (NN) term is defined -- by using the displacement vectors -- by}
\begin{equation} \label{H_NN}
H\sub{NN}  =  t_1 \sum_{j = 1}^3  (T_{\ve{d}_j}  + T_{- \ve{d}_j})\quad \text{with $t_1\in \R$}.  
\end{equation}
{The next nearest neighbor (NNN) term uses instead the periodicity vectors  and reads}
\begin{equation} \label{H_NNN}
H\sub{NNN}  = 
t_2 (\cos \phi) \sum_{j=1}^3 (T_{\ve{a}_j} + T_{- \ve{a}_j}) \,+\,
t_2 (\iu \sin \phi) (\chi_A - \chi_B) \sum_{j=1}^3 (T_{\ve{a}_j} - T_{- \ve{a}_j})  
\end{equation}
{with $t_2\in \R$}. The last term is a potential that distinguishes sites in sublattices {$\Gamma_A$ and $\Gamma_B$}, namely
\begin{equation} \label{H_site}
V_\x = M (\chi_{A} - \chi_{B})_\x =  
\begin{cases}
+ M               & \text{ if } \x \in {\Gamma_A} \\
- M                & \text{ if } \x \in {\Gamma_B}.
\end{cases}
\end{equation}

\begin{remark}[{Comparison with the honeycomb Hofstadter model}]
\label{Rem:simpler}
By analogy with the Hofstadter model \cite{Hofstadter76}, one might be tempted to replace 
the NNN term by the more symmetric expression
\begin{align} \label{tilde_H_NNN}
\widetilde H\sub{NNN}  
&= {t_2} \,\, \sum_{j=1}^3 (\eu^{\iu \phi} T_{\ve{a}_j} +  \eu^{-\iu \phi} T_{- \ve{a}_j}) \\
&= t_2 (\cos \phi) \sum_{j=1}^3 (T_{\ve{a}_j} + T_{- \ve{a}_j})  
+ t_2 (\iu \sin \phi) \sum_{j=1}^3 (T_{\ve{a}_j} - T_{- \ve{a}_j}).
\nonumber  
\end{align} 
Notice, however, that the latter operator does not distinguish between the sublattices {$\Gamma_A$ and $\Gamma_B$}, yielding
an operator which acts diagonally on the $\C^2$-factor in $\ell^2(\Gamma) \otimes \C^2$. 
The operator \eqref{H_NNN} acts instead in a non-diagonal way, and offers the opportunity to model subtler 
physical effects.  \hfill $\diamond$
\end{remark}

\noindent One can easily check that the Haldane model enjoys some relevant symmetries: 
\renewcommand{\labelenumi}{{\rm(\roman{enumi})}}
\begin{enumerate} 
\item  {$\Gamma$-periodicity}:  indeed, one checks that $[T_\gamma, H] =0$ for every $\gamma \in \Gamma$;             
\item  {$\frac{2\pi}{3}$-rotation symmetry}:  indeed  $[U_R, H] =0$ where $U_R$ is defined as usual by 
$(U_R\psi)_{\x} = \psi_{R^{-1}\x}$, with $R \in \mathrm{SO}(2)$ a rotation by a $\frac{2\pi}{3}$ angle in the plane; 
\item  {broken time-reversal symmetry (TRS)}:  for $\phi \in {\{0,\pi\}}$ the Hamiltonian commutes with the time-reversal operator, 
given by complex conjugation in $\ell^2(\cry)$. As far as $\sin \phi \neq 0$, TRS is broken, as it clearly appears from 
\eqref{H_NNN}.  
\end{enumerate}


\subsection{The Fourier decomposition}
\label{Sec:Fourier}
The $\Gamma$-periodicity of the model allows to use Fourier transform or, more intrinsically, the Bloch-Floquet decomposition.

Since the Fourier transform unitarily maps $\ell^2(\Z^d)$ into $L^2(\T^d)$, after a choice of dimerization one 
obtains an isomorphism  $\ell^2(\cry) \iso \ell^2(\Gamma, \C^2) \iso L^2(\T^2_*, \C^2)$ where the torus $\T^2_* = \R^2/\Gamma^*$, called \emph{Brillouin torus} by physicists,  is defined as a quotient by the reciprocal or dual lattice
\begin{equation} \label{Def:Reciprocal}
\Gamma^* = \set{ \ve k \in {\R^2} :  \ve k \cdot \ve \gamma \in 2\pi\Z \text{ for all } \gamma \in \Gamma}. 
\end{equation}
We choose any dimerization such that the sublattices are identified with $\Gamma$ and 
$\Gamma + \ve \nu$, respectively, for a suitable $\ve \nu \in \set{\ve d_1, \ve d_2, \ve d_3}$ (compare Figure \ref{fig:unitcells}).  
With this convention, an isomorphism is exhibited by
\begin{equation} \label{Def:Fourier}
(\F_\nu \psi)(\ve{k}) =  \sum_{\ve \gamma \in \Gamma} \eu^{- \iu \ve{k} \cdot \ve\gamma}  \Psi_{- \ve \gamma}  
=    \sum_{\ve \gamma \in \Gamma} \eu^{- \iu \ve{k} \cdot \ve\gamma}  
\left(
        \begin{array}{c}
        \psi_{- \ve \gamma}   \\
        \psi_{- \ve \gamma + \ve \nu}   
        \end{array}
\right).
\end{equation}
The operator $\F_\nu$ establishes a unitary transformation
\begin{equation}
\label{eqn:defn Fnu}
\F_\nu\colon\ell^2(\cry)\to\mathcal{H}:=
L^2(\T^2_*,\C^2), 
\end{equation}
where $\mathcal{H}$ is equipped with the inner product (notice the normalization)
$$
\inner{\varphi_1}{\varphi_2}_{\mathcal{H}}:=\frac{1}{|\T^2_*|}\int_{\T^2_*}\di \ve{k}\,
\inner{\varphi_1(\ve{k})}{\varphi_2(\ve{k})}_{\C^2}.
$$

Every operator $A$ acting in $\ell^2(\cry)$ which is $\Gamma$-periodic, in the sense that
\begin{equation} \label{Periodicity}
[A, T_{\gamma}] =0 \text{ for every } \gamma \in \Gamma,
\end{equation}
is conjugated to an operator
 $\F_{\nu} \, A \, \F_{\nu}^{-1} =: A_{\nu}$ acting in $\mathcal{H}$. Notice that $A_{\nu}$ is decomposable
\footnote{{For the sake of brevity we omit the dependence of the operator $A_{\nu}$ on the dimerization procedure, \ie we remove the subscript $\nu$.}
}  
in the sense that 
$$
(A_\nu \ph) (\ve k) = A(\ve k) \ph(\ve k)   \qquad \text{ for all } \ve k \in \T^2_* \, ,   
$$
where $\T^2_*\ni \ve k \mapsto A(\ve k)\in \mathcal{B}(\C^2)$ due to \eqref{eqn:defn Fnu}.  
Moreover, the $\Gamma$-periodicity of $A$  reflects in the following property: 
\begin{equation} \label{A periodic}
A(\ve{k} + \la) = A(\ve{k})   \qquad  \text{ for all } \la \in \Gamma^*, \ve{k} \in \T^2_{*}. 
\end{equation}
\noindent The latter is understood as an equality of matrices.  The matrix 
$A(\ve k)$ is called the \emph{fiber of the operator $A$}  at the point $\ve k$, and we use 
the notation  $A \longleftrightarrow A(\ve k)$ to indicate the correspondence between the $\Gamma$-periodic 
operator $A$  and the operator $\F_{\nu} \, A \, \F_{\nu}^{-1}$, acting in $L^2(\T^2_*, \C^2)$, 
given {fiberwise} by the multiplication operator times the  (matrix-valued) function $A(\ve{k})$. 
Notice that everything above depends -- in general -- on the choice of a dimerization, as the subscript in $\F_{\nu}$ suggests.

\bigskip

The Haldane Hamiltonian \eqref{Haldane} is $\Gamma$-periodic, and its fibers  $H(\ve{k})$ over the Brillouin torus
can be conveniently decomposed on the Pauli basis $\set{\sigma_0 = \id, \sigma_1, \sigma_2, \sigma_3}$
as  
$$
H(\ve{k}) = \sum_{j=0}^3  R_j(\ve{k}) \, \sigma_j.
$$
It is easy to show that 
\begin{eqnarray} \label{R0} 
R_0(\ve{k})  &=&  2 t_2 (\cos {\phi}) \sum_{j=1}^3 \cos(\ve{k} \cdot \ve{a}_j), \\
R_3(\ve{k})  &=&  M - 2 t_2 (\sin {\phi})  \sum_{j=1}^3 \sin(\ve{k} \cdot \ve{a}_j). 
\label{R3}
\end{eqnarray}
Indeed, one exploits the fact that the Fourier transform intertwines the translation operator 
$T_{\gamma}$, for $\gamma \in \Gamma$, with the multiplication times $\eu^{\iu \ve{k} \cdot \gamma} \, \id$.  
Since $T_{\ve{a}_j}  \longleftrightarrow   \eu^{\iu \ve{k} \cdot \ve{a}_j} \id$, one concludes that
$$ 
T_{\ve{a}_j} + T_{-\ve{a}_j}  \longleftrightarrow   2 \cos(\ve{k} \cdot \ve{a}_j) \, \id     
$$
which immediately gives \eqref{R0}.  Analogously, since
$
(\chi_A - \chi_B) T_{\ve{a}_j}  \longleftrightarrow   \eu^{\iu \ve{k} \cdot \ve{a}_j } \, \sigma_3     
$,
one concludes that 
$$ 
(\chi_A - \chi_B) \left(T_{\ve{a}_j} - T_{- \ve{a}_j} \right) \longleftrightarrow 2\iu \sin(\ve k \cdot \ve a_j) \, \sigma_3     
$$
which gives \eqref{R3}. Notice that the previous terms do not depend on a specific choice of the dimerization, provided one 
of the sublattices agrees with $\Gamma$. 

As for the off-diagonal terms, one has however to be more careful, since the computation \crucial{does depend on the choice of ``the'' periodicity cell}, as pointed out for example in \cite{BenaMontambaux,Fruchart2014}. 
We make here the specific choice 
\begin{equation} \label{cellY}
Y = \set{\ve{x} \in \R^2:  \ve{x} = \alpha_1 \ve a_1 + \alpha_2 \ve a_2 \text{ with } \alpha_j \in [-\half,+\half] } 
\end{equation}
so that $Y \cap \, \cry = \set{\ve 0, \ve d_3}$, as illustrated in the first panel in Figure 2. 
One has that $\cry \iso \Gamma \times \set{0, \ve{d}_3}$ as a measure space, and the dimerization isomorphism is exhibited by
\begin{equation} \label{Specific_dimer}
\Psi_{\gamma} = 
\left(
\begin{array}{c}
\psi_{\gamma, A}   \\
\psi_{\gamma, B}   
\end{array}
\right)
= 
\left(
\begin{array}{c}
\psi_{\gamma + 0}  \\
\psi_{\gamma + \ve{d}_3}   
\end{array}
\right) \, .
\end{equation}

\noindent With this specific choice, the remaining terms are 
\begin{eqnarray} \label{R1}
R_1(\ve{k})  &=&  t_1  \(  1 + \cos(\ve{k} \cdot  \ve{a}_1)  +  \cos(\ve{k} \cdot \ve{a}_2) \), \\
R_2(\ve{k})  &=&  t_1 \(  \sin(\ve{k} \cdot  \ve{a}_1)  - \sin(\ve{k} \cdot \ve{a}_2) \).  
\label{R2}
\end{eqnarray}

\noindent These expressions are easily derived.  
By using \eqref{Translations}, one computes
\begin{eqnarray*}
\( T_{ +\ve{d}_3} \psi  \)_{\gamma} = 
\left(
\begin{array}{c}
\psi_{\gamma - \ve{d}_3}   \\
\psi_{(\gamma + \ve{d}_3) - \ve{d}_3}   
\end{array}
\right)
=
\left(
\begin{array}{c}
0  \\
\psi_{\gamma, A}   
\end{array}
\right)  ,  \\
\( T_{- \ve{d}_3} \Psi  \)_{\gamma} = 
\left(
\begin{array}{c}
\psi_{\gamma + \ve{d}_3}   \\
\psi_{(\gamma + \ve{d}_3) + \ve{d}_3}   
\end{array}
\right)
=
\left(
\begin{array}{c}
\psi_{\gamma, B} \\
0
\end{array}
\right).   \\
\end{eqnarray*}
Thus $T_{\ve{d}_3}  + T_{- \ve{d}_3}  = \Id \otimes \sigma_1$, so that the Fourier transform $\F_{\ve d_3}$ yields
\begin{equation} \label{}
T_{\ve{d}_3}  + T_{- \ve{d}_3} \longleftrightarrow  \Id \otimes \sigma_1  .
\end{equation}
The coordinate $j=3$ is privileged in view of our choice of the periodicity cell.
  
As for the next term, one uses that $\ve{a}_1 = \ve{d}_2 - \ve{d}_3$ so that
\begin{eqnarray*}
\( T_{ +\ve{d}_2} \psi  \)_{\gamma} = 
\left(
\begin{array}{c}
\psi_{\gamma - \ve{d}_3 - \ve a_1}   \\
\psi_{(\gamma + \ve{d}_3 ) - \ve{d}_3  - \ve{a}_1}   
\end{array}
\right)
=
\left(
\begin{array}{c}
0  \\
\psi_{\gamma - \ve{a}_1, A}   
\end{array}
\right) , \\
\( T_{- \ve{d}_2} \psi  \)_{\gamma} = 
\left(
\begin{array}{c}
\psi_{\gamma + \ve{d}_3 + \ve{a}_1}   \\
\psi_{(\gamma + \ve{d}_3) + \ve{d}_3 + \ve a_1}   
\end{array}
\right)
=
\left(
\begin{array}{c}
\psi_{\gamma + \ve{a}_1, B} \\
0
\end{array}
\right).   \\
\end{eqnarray*}
After Fourier transform one obtains 
\begin{eqnarray*}
\(\F_{\ve d_3} (T_{\ve{d}_2}  + T_{- \ve{d}_2})\psi \) (\ve{k}) &=&
\left(
\begin{array}{cc}
 0    &   \eu^{{-} \iu \ve{k} \cdot \ve{a}_1} \\
  \eu^{{+}\iu \ve{k} \cdot \ve{a}_1}    &   0
\end{array}
\right)
\(\F_{\ve d_3} \psi \) (\ve{k}). 
\end{eqnarray*}

\noindent Analogously, in view of  $\ve{a}_2 = \ve{d}_3 - \ve{d}_1$ one has
\begin{eqnarray*}
\( T_{ +\ve{d}_1} \psi  \)_{\gamma} = 
\left(
\begin{array}{c}
\psi_{\gamma - \ve{d}_3 + \ve a_2}   \\
\psi_{(\gamma + \ve{d}_3)  - \ve{d}_3 + \ve a_2}   
\end{array}
\right)
=
\left(
\begin{array}{c}
0  \\
\psi_{\gamma + \ve{a}_2, A}   
\end{array}
\right) ,  \\
\( T_{- \ve{d}_1} \psi  \)_{\gamma} = 
\left(
\begin{array}{c}
\psi_{\gamma + \ve{d}_3 -\ve a_2}   \\
\psi_{(\gamma + \ve{d}_3) +\ve{d}_3 -\ve a_2}   
\end{array}
\right)
=
\left(
\begin{array}{c}
\psi_{\gamma - \ve{a}_2, B} \\
0
\end{array}
\right) ,  \\
\end{eqnarray*}
which gives 
\begin{eqnarray*}
\(\F_{\ve d_3}  (T_{\ve{d}_1}  + T_{- \ve{d}_1})\psi \) (\ve{k}) &=&
\left(
\begin{array}{cc}
 0    &   \eu^{+ \iu \ve{k} \cdot \ve{a}_2}\\
  \eu^{- \iu \ve{k} \cdot \ve{a}_2}    &   0
\end{array}
\right)
\(\F_{\ve d_3} \psi \) (\ve{k}). 
\end{eqnarray*}
Summarizing the information above, one concludes that  
\begin{equation} \label{dm1}
\sum_{j = 1}^3  (T_{\ve{d}_j}  + T_{- \ve{d}_j})  
\quad \longleftrightarrow  \quad 
\left(
\begin{array}{cc}
 0    &  1 +  \eu^{ {-} \iu \ve{k} \cdot \ve{a}_1} + \eu^{+ \iu \ve{k} \cdot \ve{a}_2}\\
1 + \eu^{{+}\iu \ve{k} \cdot \ve{a}_1}  + \eu^{- \iu \ve{k} \cdot \ve{a}_2}  &   0
\end{array} 
\right) 
\end{equation}
which immediately gives \eqref{R1} and \eqref{R2}.

\begin{remark} \label{Rmk:comparison}
Our definition of the Haldane model agrees with the ones in the cited references \cite{FruchartCarpentier2013, 
GiulianiMastropietroPorta2017, Santoro_LN},  up to the translation to first-quantization formalism and some trivial relabelling. 

\hfill $\diamond$
\end{remark}

\goodbreak

\section{Bloch functions and their singularities}
\label{Sec:Singularities}

In this Section, we will be interested in studying the spectral properties of the Haldane Hamiltonian, which we rewrite as
\[
H(\ve{k}) = \sum_{j=0}^{3} R_j(\ve{k}) \, \sigma_j = 
\begin{pmatrix}
R_0(\ve{k}) + R_3(\ve{k}) & \overline{R(\ve{k})} \\[3pt]
R(\ve{k}) & R_0(\ve{k}) - R_3(\ve{k})
\end{pmatrix},
\]
where we have abbreviated 
\[ R(\ve{k}) := R_1(\ve{k}) + \iu\, R_2(\ve{k}) = t_1 \big(1 +  \eu^{\iu \ve{k} \cdot \ve{a}_1} + \eu^{-\iu \ve{k} \cdot \ve{a}_2}\big)\] 
(compare \eqref{dm1}). It is then immediate to see that the eigenvalues of $H(\ve{k})$ are given by
\[ E_{\pm}(\ve{k}) := R_0(\ve{k}) \pm \sqrt{\sum_{j=1}^{3} R_j(\ve{k})^2} = R_0(\ve{k}) \pm \sqrt{R_3(\ve{k})^2 + \left| R(\ve{k}) \right|^2}. \]
These two energy bands will not  overlap (that is, $E_-(\ve{k}) \le E_+(\ve{k})$ for all $\ve{k} \in \R^2$) provided that $t_1 \neq 0$.  For simplicity, in the following we will assume $t_1,t_2 >0$. The bands can still touch at the points in the Brillouin torus which are determined by the equation
\[ \sum_{j=1}^{3} R_j(\ve{k})^2 = 0 \quad \Longleftrightarrow \quad R(\ve{k}) = 0 \text{ and } R_3(\ve{k}) = 0. \]
We see then that there are (at most) two such points in the Brillouin torus, usually labeled $\ve{K}$ and $\ve{K}'$, determined by the zeroes of $R$: these are obtained from the conditions 
\[ \eu^{\iu \ve{K}' \cdot \ve{a}_1} = \eu^{\iu 2 \pi /3} \text{ and } \eu^{-\iu \ve{K}' \cdot \ve{a}_2} = \eu^{-\iu 2 \pi /3}, \quad \eu^{\iu \ve{K} \cdot \ve{a}_1} = \eu^{-\iu 2 \pi /3} \text{ and } \eu^{-\iu \ve{K} \cdot \ve{a}_2} = \eu^{\iu 2 \pi /3}, \]
which in particular imply $\ve{K}'=-\ve{K} \bmod \Gamma^*$. 
Since locally around these points the dispersion of the energy bands is linear when they produce band intersections, \ie $E_{\pm}(\ve{k})=E_{\pm}(\ve{K}^{\sharp})\pm v\sub{F} | \ve{k}-\ve{K}^{\sharp}| + O\big(| \ve{k}-\ve{K}^{\sharp}|^2\big)$ for $\ve{K}^{\sharp} \in \set{\ve{K},\ve{K}'}$, the points $\ve{K}$ and $\ve{K}'$ are usually called \emph{Dirac points}. 
The equation $R_3(\ve{k})=0$ then determines the locus in the space of parameters $(\phi,M)$ where either $\ve{K}$ or $\ve{K}'$ (or both) are points of degeneracy for the eigenvalues of the Haldane Hamiltonian, namely%
\footnote{
Notice that when $R(\ve{k})=0$, that is, when $R_1(\ve{k})=t_1 \big( 1+\cos(\ve{k} \cdot \ve{a}_1) + \cos(\ve{k} \cdot \ve{a}_2) \big) = 0$ and $R_2(\ve{k})= t_1 \big(\sin(\ve{k} \cdot \ve{a}_1) - \sin(\ve{k} \cdot \ve{a}_2)\big) =0$, then using $\ve{a}_3 = - \ve{a}_1 - \ve{a}_2$
\begin{align*}
\sin(\ve{k} \cdot \ve{a}_3) &= - \sin(\ve{k} \cdot \ve{a}_1 + \ve{k} \cdot \ve{a}_2) = - \big( \sin(\ve{k} \cdot \ve{a}_1) \, \cos(\ve{k} \cdot \ve{a}_2) + \sin(\ve{k} \cdot \ve{a}_2) \, \cos(\ve{k} \cdot \ve{a}_1) \big) \\
& = - \big( \sin(\ve{k} \cdot \ve{a}_1) \,(-1 -\cos(\ve{k} \cdot \ve{a}_1) )+ \sin(\ve{k} \cdot \ve{a}_1) \, \cos(\ve{k} \cdot \ve{a}_1) \big) \\
& = \sin(\ve{k} \cdot \ve{a}_1),
\end{align*}
so that $\sum_{j=1}^{3} \sin(\ve{k} \cdot \ve{a}_j) = 3 \, \sin(\ve{k}\cdot\ve{a}_1)$. Now $\sin(\ve{K}\cdot\ve{a}_1) = \sin(-2\pi/3) = - \sqrt{3}/2$, while $\sin(\ve{K}'\cdot\ve{a}_1) = \sin(2\pi/3) = \sqrt{3}/2$.
}
\[ R_3(\ve{K}) = M + 3 \sqrt{3}\, t_2 \, \sin \phi, \quad R_3(\ve{K}') = M - 3 \sqrt{3} \, t_2 \, \sin \phi. \]

\begin{figure}[htb]
\centering
\includegraphics[width=.8\textwidth]{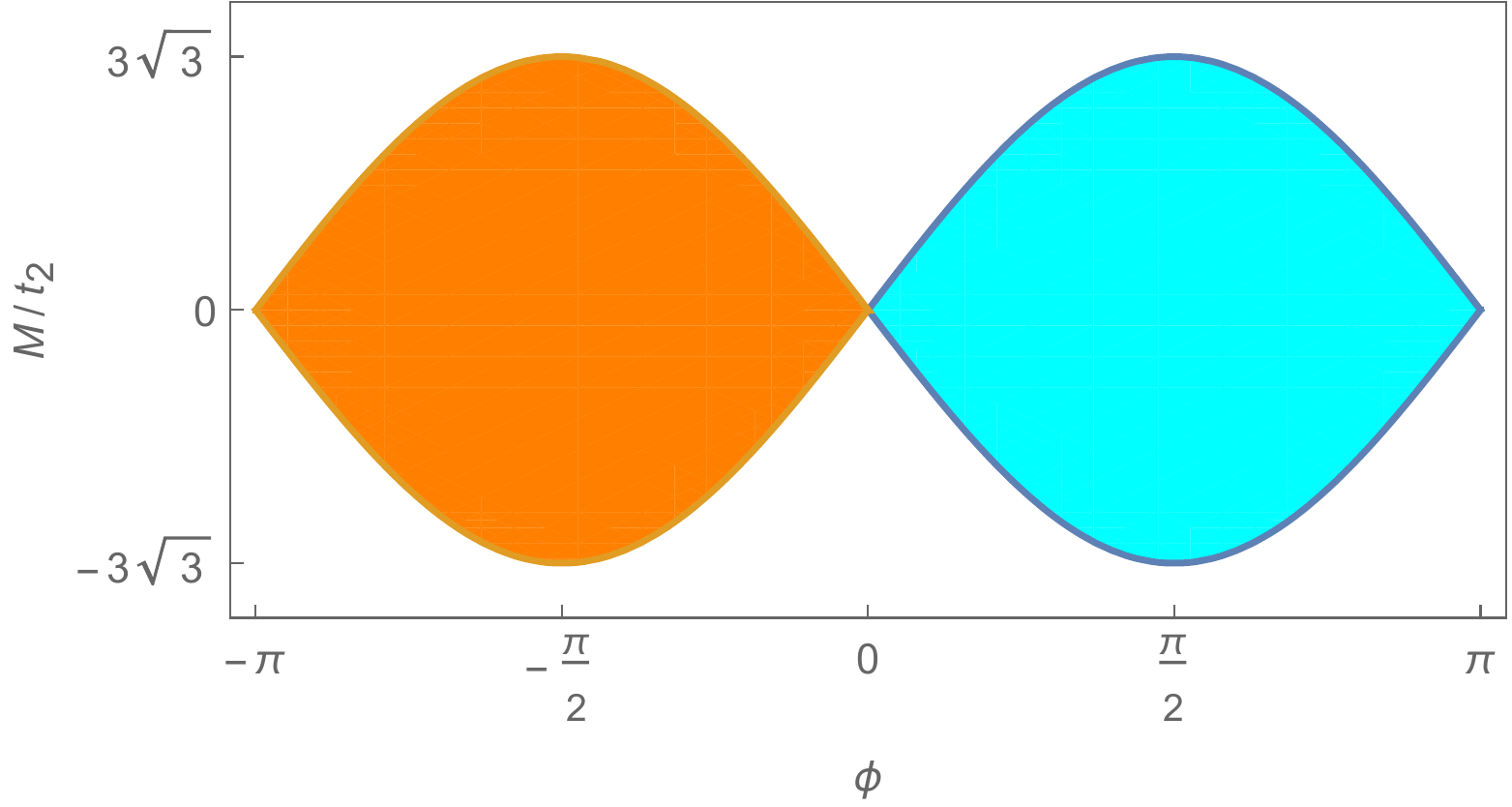}
\caption{\footnotesize The topological phase diagram of the Haldane model. In cyan, the region $\set{R_3(\ve{K}) > 0,  R_3(\ve{K}') < 0}$, characterized by a Chern number $c_1=-1$; in orange, the region $\set{R_3(\ve{K}) < 0,  R_3(\ve{K}') > 0}$, characterized by a Chern number $c_1=+1$ (color online). In the rest of the phase diagram, $c_1=0$.} 
\label{PhaseDiagram}
\end{figure}

We see that the parameter space $(\phi,M)$ gets divided into four regions where the Hamiltonian is gapped (see Figure~\ref{PhaseDiagram}), characterized by the signs of $R_3(\ve{K})$ and $R_3(\ve{K}')$. We will show now how it is possible to assign a \emph{topological label} to each of the four gapped phases, determining also the ``quantum anomalous Hall conductivity'' of the Haldane model for all parameters in the region. To this end, it is convenient to introduce the eigenvector $u_-(\ve{k})$, that is, the \emph{Bloch function}, associated to the lower band $E_-(\ve{k})$ of the Haldane Hamiltonian. This reads
\[ u_-(\ve{k}) = N(\ve{k})^{-1} \,
\begin{pmatrix}  
\sqrt{R_3(\ve{k})^2 + |R(\ve{k})|^2} - R_3(\ve{k}) \\
-R(\ve{k})
\end{pmatrix}, \]
where $N(\ve{k}) := \left[ 2 \sqrt{R_3(\ve{k})^2 + |R(\ve{k})|^2} \left( \sqrt{R_3(\ve{k})^2 + |R(\ve{k})|^2} - R_3(\ve{k}) \right) \right]^{1/2}$ is a normalizing factor ensuring $\norm{u_-(\ve{k})}_{\C^2} = 1$ for all $\ve{k} \in \R^2$ (compare \cite[Appendix B]{GiulianiMastropietroPorta2017}). The \emph{Bloch gauge} (that is, the phase within the complex one-dimensional eigenspace associated to the lower energy band) is chosen so that the first component $u_{-,1}(\ve{k})$ is real.

\begin{figure}[ht]
\centering
\begin{subfigure}{.3\textwidth}
\centering
\includegraphics[width=\textwidth]{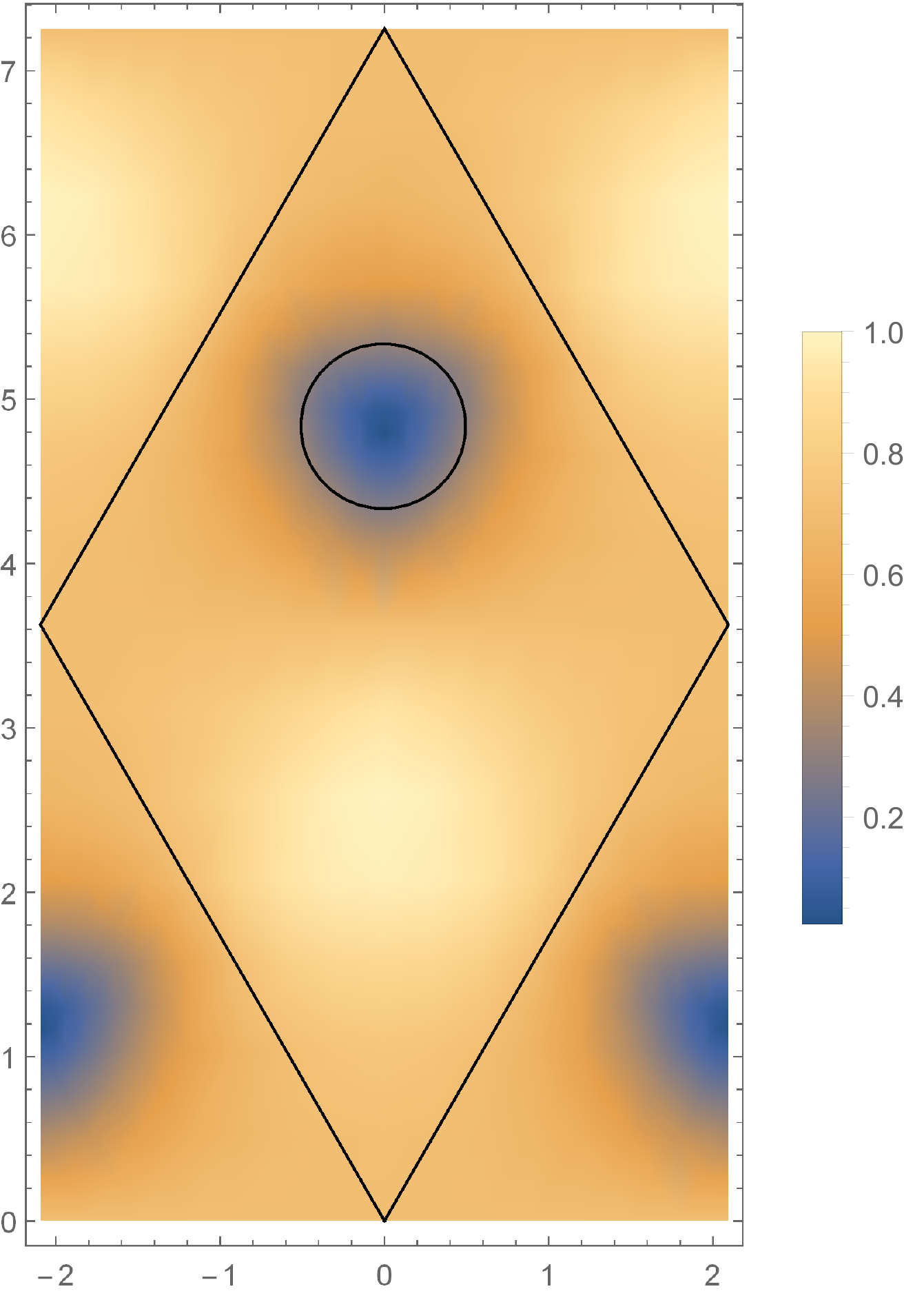}
\caption{$u_{-,1}(\ve{k})$}
\label{subfig:A}
\end{subfigure}
\begin{subfigure}{.3\textwidth}
\centering
\includegraphics[width=\textwidth]{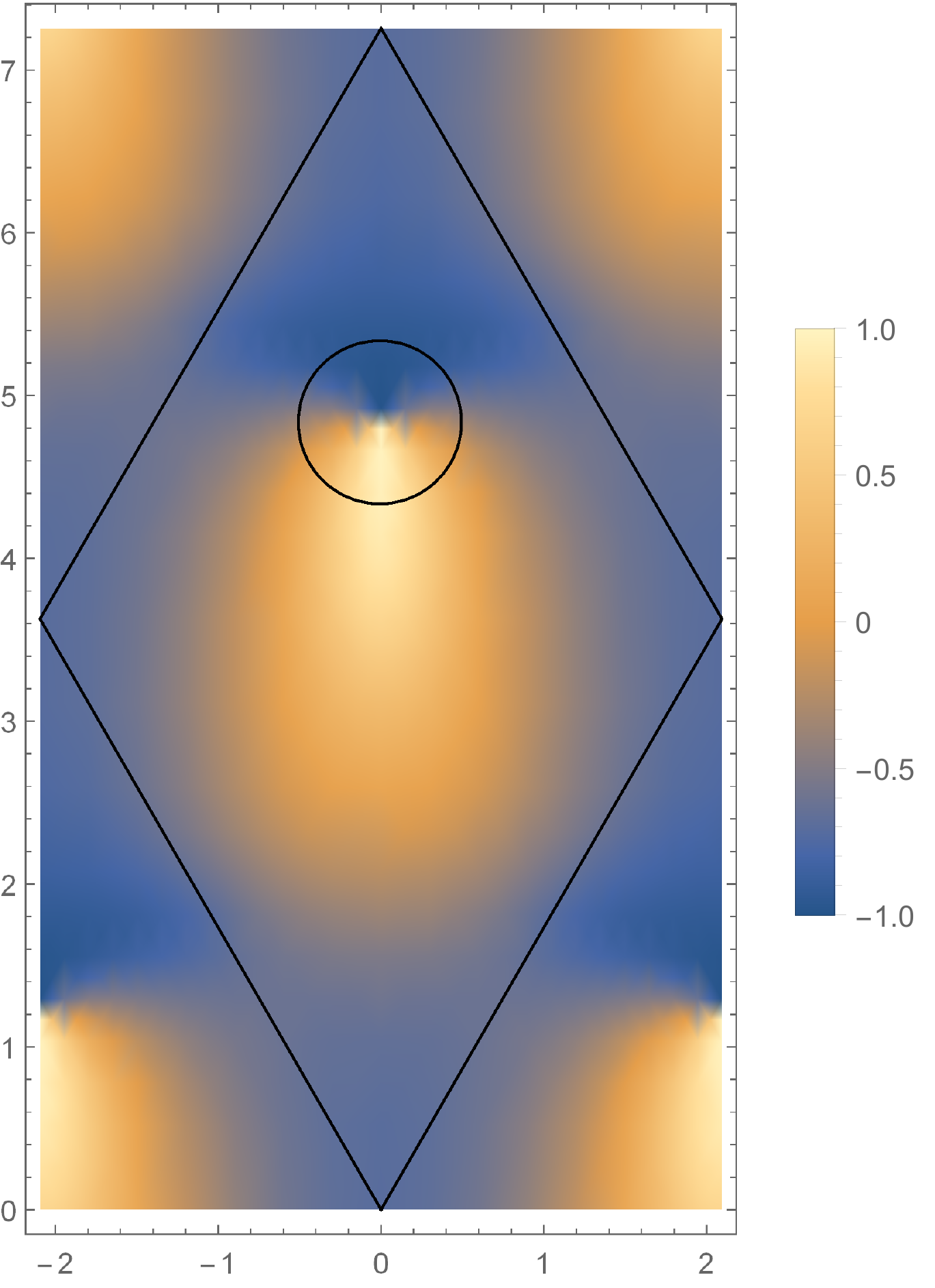}
\caption{$\operatorname{Re} u_{-,2}(\ve{k})$}
\label{subfig:B}
\end{subfigure}
\begin{subfigure}{.3\textwidth}
\centering
\includegraphics[width=\textwidth]{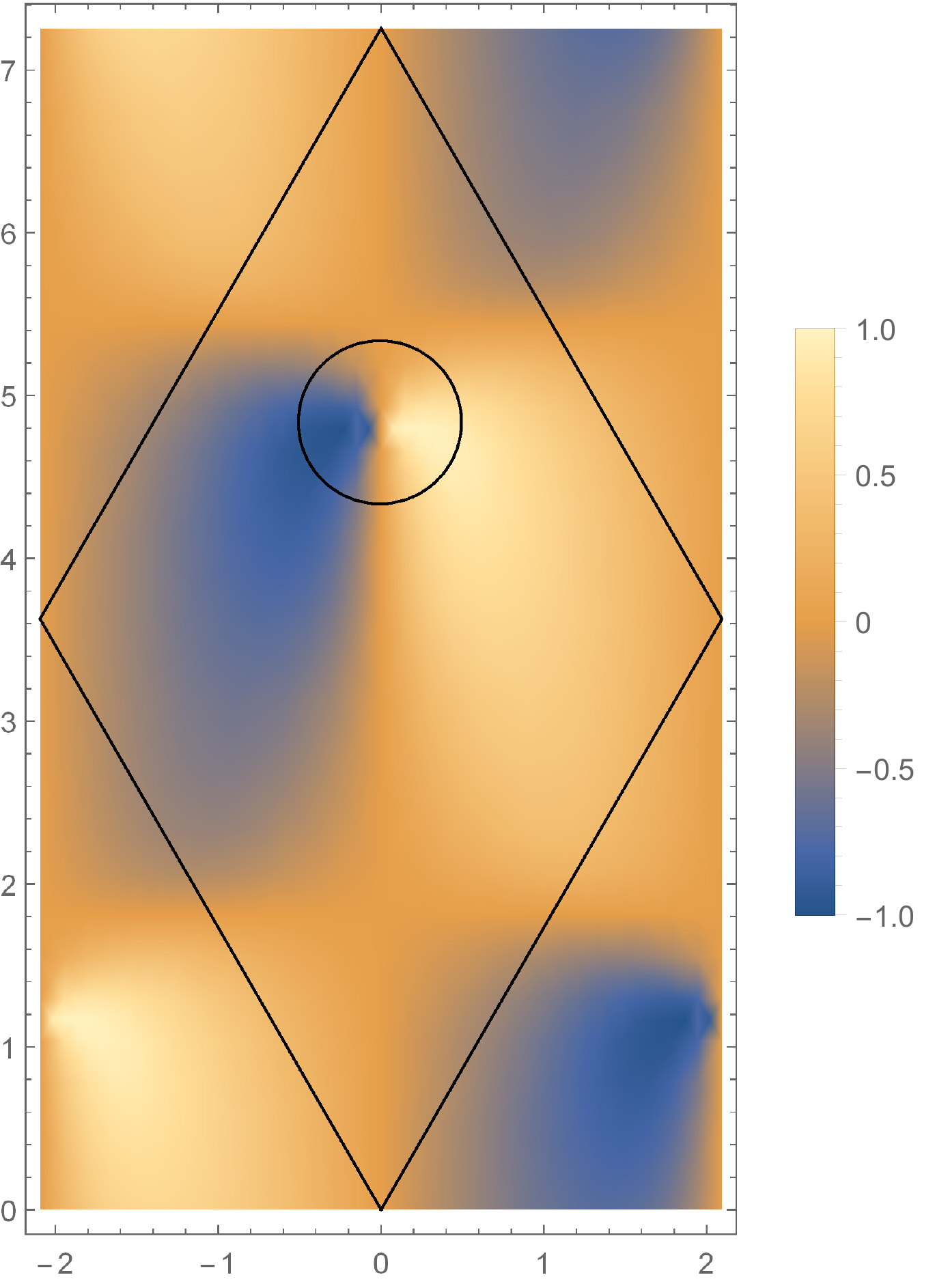}
\caption{$\operatorname{Im} u_{-,2}(\ve{k})$}
\label{subfig:C}
\end{subfigure}

\begin{minipage}[c]{.3\textwidth}
\centering
\vspace{0pt}
\begin{subfigure}{\textwidth}
\centering
\includegraphics[width=\textwidth]{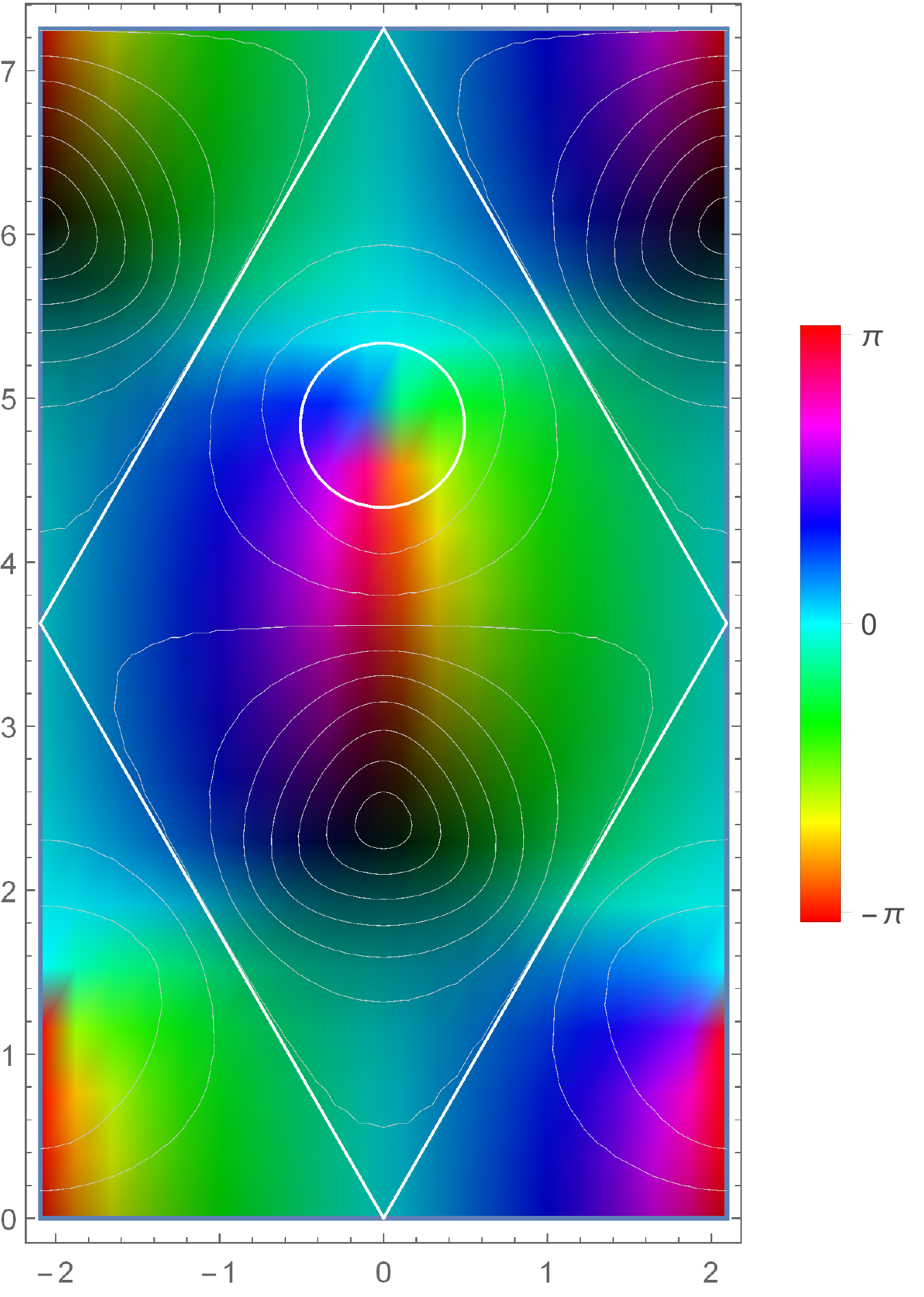}
\caption{$|u_{-,2}(\ve{k})|$}
\label{subfig:D}
\end{subfigure}
\end{minipage}
\begin{minipage}[c]{.6\textwidth}
\centering
\vspace{0pt}
\captionsetup{width=.9\textwidth}
\caption{Density plots for the components of $u_-(\ve{k})$ (color online). The parameters chosen to produce these plots are as follows: $d=1$ for the lattice constant, $t_1=1$, $t_2=1/4$, $M=0$, $\phi = \pi/2$. The rhomboidal region is the Brillouin zone $\set{k_1 \, \ve{b}_1 + k_2 \, \ve{b}_2: k_1,k_2 \in [0,1]}$, where the vectors $\ve{b}_1,\ve{b}_2$ spanning the dual lattice $\Gamma^*$ are determined by the conditions $\ve{a}_i \cdot \ve{b}_j = 2 \pi \delta_{ij}$. The circle points to the position of the Dirac point $\ve{K}$. The rapid change of both $\operatorname{Re} u_{-,2}(\ve{k})$ and $\operatorname{Im} u_{-,2}(\ve{k})$ around $\ve{k}=\ve{K}$ are evident from \eqref{subfig:B} and \eqref{subfig:C}, respectively, signalling a discontinuity. Instead, $u_{-,1}(\ve{k})$ is seen to be regular (and actually vanishing) at $\ve{k}=\ve{K}$ from \eqref{subfig:A}. In the last plot \eqref{subfig:D}, contour lines for the absolute value $|u_{-,2}(\ve{k})|$ are plotted, while the color code indicates the value of the argument of the phase of $u_{-,2}(\ve{k})$ as in the legend. In agreement with the previous comments, it is possible to see a phase singularity of $u_{-,2}(\ve{k})$ around $\ve{k}=\ve{K}$, while the absolute value $|u_{-,2}(\ve{k})| = \sqrt{1-u_{-,1}(\ve{k})^2}$ remains smooth.}
\label{fig:phi!=0}
\end{minipage}
\end{figure}

\begin{figure}[ht]
\centering
\begin{subfigure}{.3\textwidth}
\centering
\includegraphics[width=\textwidth]{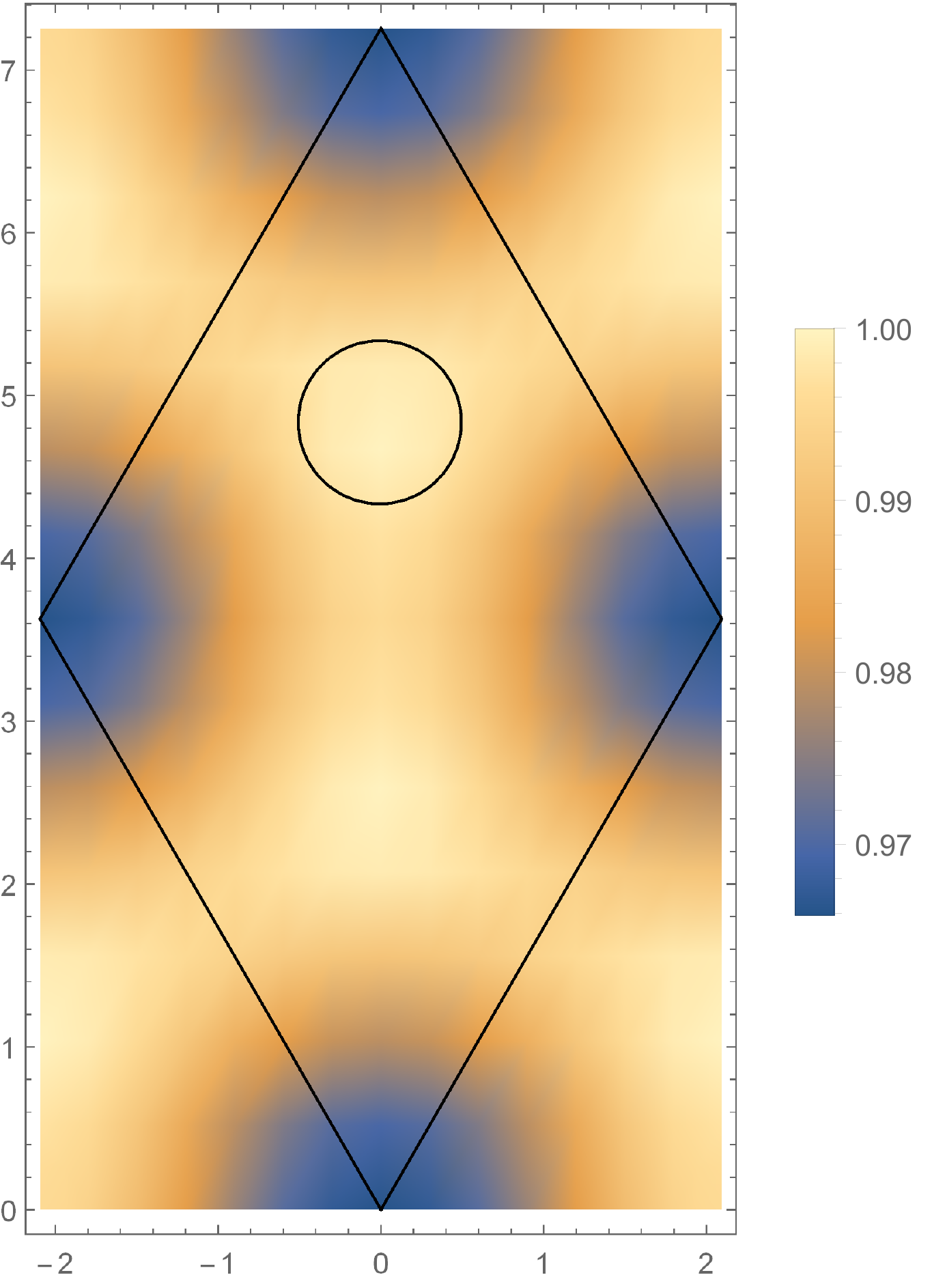}
\caption{$u_{-,1}(\ve{k})$}
\end{subfigure}
\begin{subfigure}{.3\textwidth}
\centering
\includegraphics[width=\textwidth]{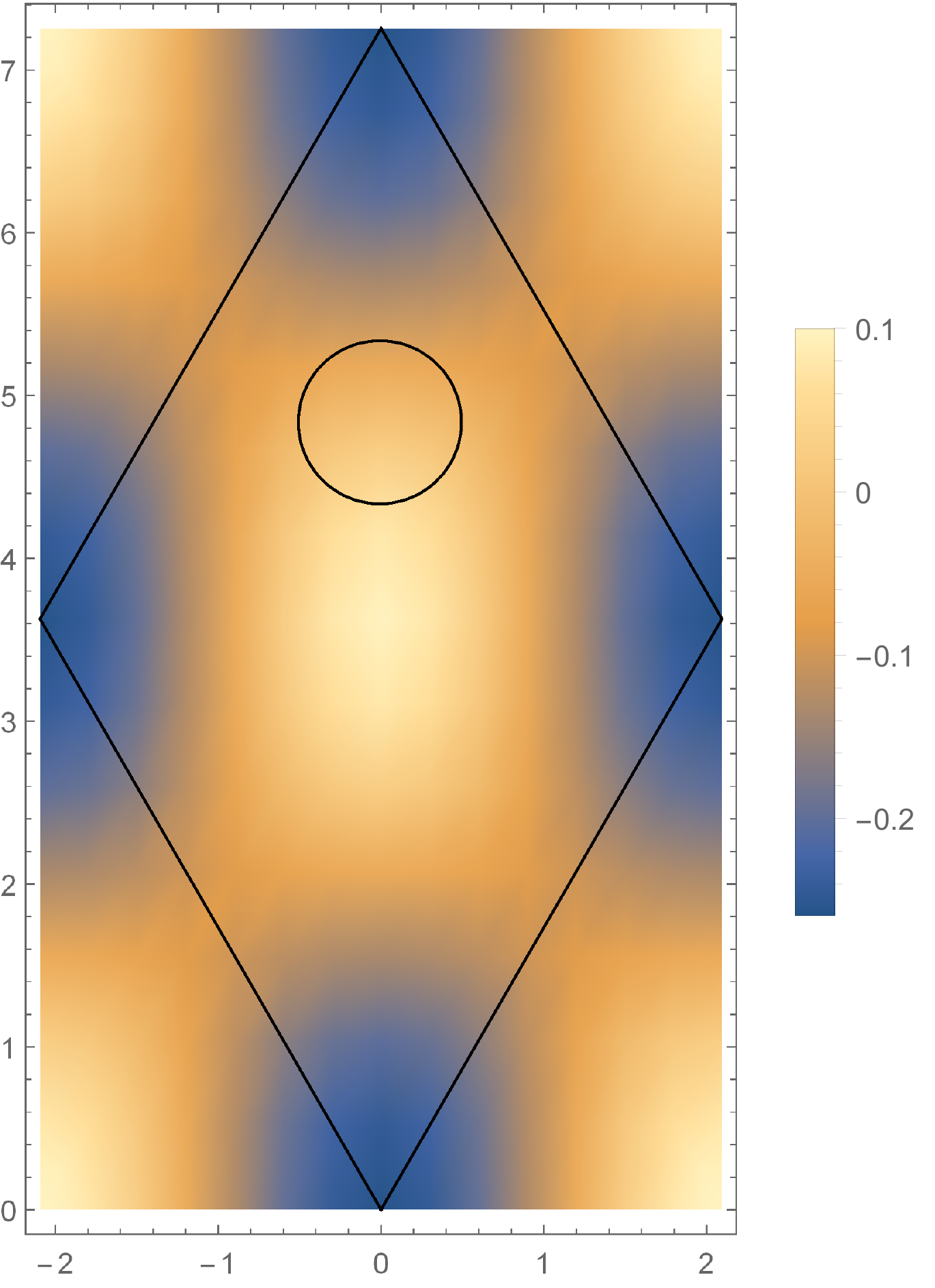}
\caption{$\operatorname{Re} u_{-,2}(\ve{k})$}
\end{subfigure}
\begin{subfigure}{.3\textwidth}
\centering
\includegraphics[width=\textwidth]{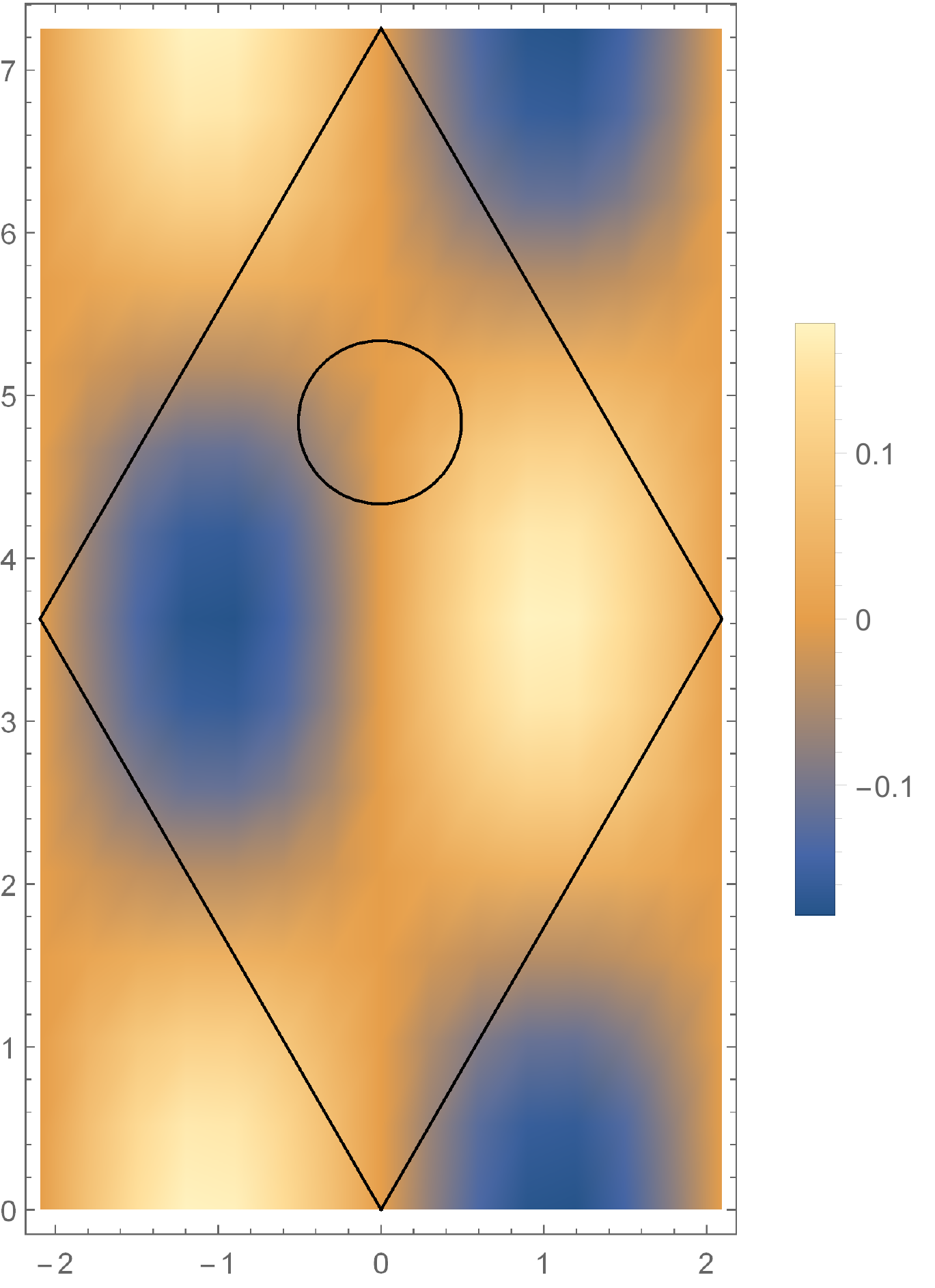}
\caption{$\operatorname{Im} u_{-,2}(\ve{k})$}
\end{subfigure}

\begin{minipage}[c]{.3\textwidth}
\centering
\vspace{0pt}
\begin{subfigure}{\textwidth}
\centering
\includegraphics[width=\textwidth]{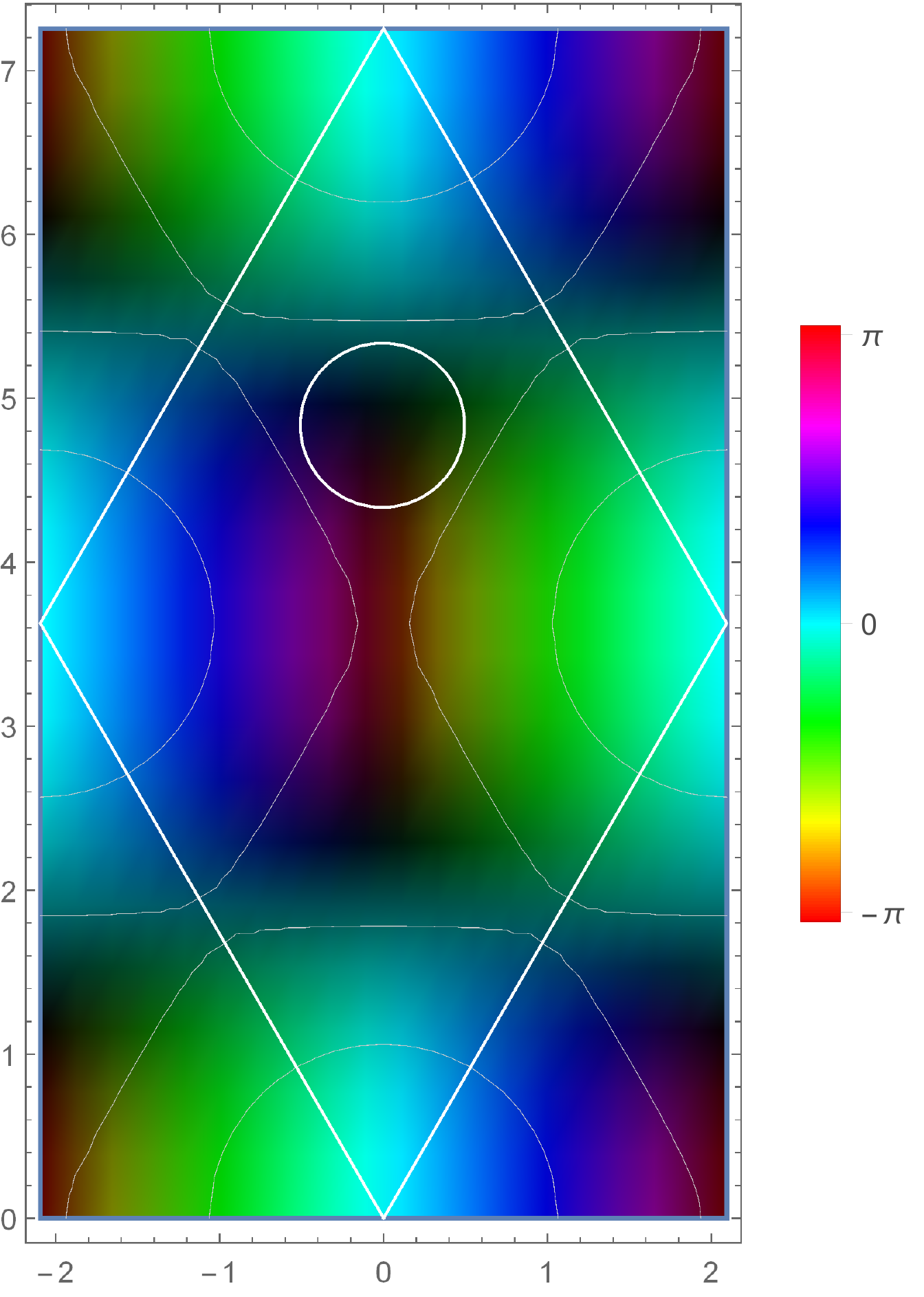}
\caption{$|u_{-,2}(\ve{k})|$}
\end{subfigure}
\end{minipage}
\begin{minipage}[c]{.6\textwidth}
\centering
\vspace{0pt}
\captionsetup{width=.9\textwidth}
\caption{Similar plots to those of Figure \ref{fig:phi!=0} (color online), this time corresponding to the parameters $M=-3\sqrt{3}$ and $\phi = 0$. All other parameters where left as specified in Figure \ref{fig:phi!=0}. In this case, $u_-(\ve{k})$ is analytic over the whole Brillouin zone.}
\label{fig:phi=0}
\end{minipage}
\end{figure}

If $\ve{K}^\sharp$ denotes either of the Dirac points, then $R(\ve{K}^\sharp)$ vanishes, as $\ve{K}$ and $\ve{K}'$ are precisely the zeroes of $R$, while $R_3(\ve{K}^\sharp) \ne 0$ due to the gap condition. Consequently, 
\[ \sqrt{R_3(\ve{k})^2 + |R(\ve{k})|^2} - R_3(\ve{k}) \Big|_{\ve{k}=\ve{K}^\sharp} = \big|R_3(\ve{K}^\sharp)\big|-R_3(\ve{K}^\sharp) , \]
and similarly
\[ N(\ve{K}^\sharp) = \left[ 2 \big|R_3(\ve{K}^\sharp)\big| \left( \big|R_3(\ve{K}^\sharp)\big| - R_3(\ve{K}^\sharp) \right) \right]^{1/2}. \]
We see that $u_-(\ve{k})$ may have singularities at the Dirac points, depending on the signs of $R_3(\ve{K})$ and $R_3(\ve{K}')$. In particular, it holds that $u_-(\ve{k})$ is \emph{singular at $\ve{K}$} in the region $\set{R_3(\ve{K})>0, R_3(\ve{K}')<0}$ of the parameter space $(\phi,M)$ (depicted in cyan in Figure \ref{PhaseDiagram}), while it is \emph{analytic on the whole Brillouin torus} in the region $\set{R_3(\ve{K})<0, R_3(\ve{K}')<0}$ (the lower white region in Figure \ref{PhaseDiagram}). The qualitative features of this singularity (or lack thereof) are illustrated in Figures \ref{fig:phi!=0} and \ref{fig:phi=0}.

To investigate further the singularity of $u_-(\ve{k})$, we restrict our attention to parameters $(\phi,M)$ so that $R_3(\ve{K})>0$ and $R_3(\ve{K}')<0$. As discussed, in this region $\ve{K}$ is the only singular point of $u_-(\ve{k})$. By rewriting, after a few simple algebraic manipulations,
\[ u_-(\ve{k}) = \frac{1}{\sqrt{2}}
\begin{pmatrix}
\dfrac{|R(\ve{k})|}{\left(R_3(\ve{k})^2 + |R(\ve{k})|^2\right)^{1/4} \left( \sqrt{R_3(\ve{k})^2 + |R(\ve{k})|^2} + R_3(\ve{k}) \right)^{1/2}} \\[3pt]
-\dfrac{R(\ve{k})}{|R(\ve{k})|} \, \left( 1 + \dfrac{R_3(\ve{k})}{\sqrt{R_3(\ve{k})^2 + |R(\ve{k})|^2}} \right)^{1/2}
\end{pmatrix}  ,
\]
we see that in this region the first component of $u_-(\ve{k})$ is smooth, while it is the second component that has a singularity, due to the explicit dependence on the phase of $R(\ve{k})$. This implies in particular that, locally around $\ve{k}=\ve{K}$, $u_{-,2}(\ve{k})$ is homogeneous of degree zero in the radial coordinate $r = |\ve{k}-\ve{K}|$, so that the derivatives of $u_{-,2}(\ve{k})$ have a $(1/r)$-singularity, making the $H^1$-norm of $u_-$
\begin{align*}
\norm{u_-}_{H^1} & := \left( \norm{u_-}_{L^2}^2 + \norm{\partial_{k_1} u_-}_{L^2}^2 + \norm{\partial_{k_2} u_-}_{L^2}^2 \right)^{1/2} \\
& = \left[ \int_{\T^2_*} \left( \norm{u_-(\ve{k})}_{\C^2}^2 + \norm{\partial_{k_1} u_-(\ve{k})}_{\C^2}^2 + \norm{\partial_{k_2} u_-(\ve{k})}_{\C^2}^2 \right) \di \ve{k}  \right]^{1/2}
\end{align*}
divergent. From the same type of homogeneity argument, one can also deduce that all the fractional Sobolev norms $\norm{u_-}_{H^s}$ for $s\in [0,1)$ are instead finite.

The singularity of $u_-(\ve{k})$ at $\ve{k} = \ve{K}$ carries also a topological information. This can be accessed by means of the \emph{Berry connection}, defined as the differential $1$-form
\[ \mathcal{A} := \operatorname{Im} \, \inner{u_-(\ve{k})}{\di u_-(\ve{k})}_{\C^2} = \sum_{j=1}^{2} \operatorname{Im} \, \inner{u_-(\ve{k})}{\partial_{k_j} u_-(\ve{k})}_{\C^2} \di k_j. \]
We argue as above: the $(1/r)$-singularity of the derivatives of $u_{-,2}(\ve{k})$ around $\ve{k}=\ve{K}$ is integrable (even though not square-integrable), so that the integral of $\mathcal{A}$ around a small loop $\ell_{\eps}$ (say, of diameter $\eps \ll 1$) encircling the singularity of $u_-(\ve{k})$ stays bounded even in the limit $\eps \to 0$. Denoting by $D_{\eps}$ the region bounded by the loop $\ell_{\eps}$ (which then bounds also $\T^2_* \setminus D_{\eps}$, as the Brillouin torus is closed), in the limit of a very small loop one obtains from Stokes' theorem
\begin{equation} \label{ChernSingularity}
\lim_{\eps \to 0} \oint_{\ell_{\eps}} \mathcal{A} = - \lim_{\eps \to 0} \oint_{\partial(\T^2_* \setminus D_{\eps})} \mathcal{A}  = - \lim_{\eps \to 0} \int_{\T^2_* \setminus D_{\eps}} \di \mathcal{A} = - \int_{\T^2_*} \mathcal{F}. 
\end{equation}
In the last step, we introduced the \emph{Berry curvature} $2$-form
\begin{equation} \label{Berry}
\begin{aligned}
\mathcal{F} & := \di \mathcal{A} = 2 \operatorname{Im} \, \inner{\partial_{k_1} u_-(\ve{k})}{\partial_{k_2} u_-(\ve{k})} \di k_1 \wedge \di k_2 \\
& = - \iu \, \Tr_{\C^2} \big( P_-(\ve{k}) \left[ \partial_{k_1} P_-(\ve{k}), \partial_{k_2} P_-(\ve{k}) \right] \big) \, \di \ve{k}.
\end{aligned}
\end{equation}
The last equality (see \eg \cite[Lemma 7.2]{MonacoPanatiPisanteTeufel2018}) shows that $\mathcal{F}$ can be expressed directly in terms of the family of projections 
\begin{align*}
& P_-(\ve{k}) := \ket{u_-(\ve{k})} \bra{u_-(\ve{k})}= \\
& \frac{1}{2 \sqrt{R_3(\ve{k})^2 + |R(\ve{k})|^2}}
\begin{pmatrix}
\sqrt{R_3(\ve{k})^2 + |R(\ve{k})|^2} - R_3(\ve{k}) & - R(\ve{k}) \\[3pt]
- \overline{R(\ve{k})} & \sqrt{R_3(\ve{k})^2 + |R(\ve{k})|^2} + R_3(\ve{k})
\end{pmatrix}
\end{align*}
on the eigenspace corresponding to the lower energy band. Contrary to the Bloch function, these projections depend analytically on $\ve{k}$ over the whole Brillouin torus, making it possible to compute the last limit in \eqref{ChernSingularity}.

The Berry curvature is a geometric object. In fact, its integral over the Brillouin torus is an integer multiple of $2\pi$: 
\begin{equation} \label{Chern}
c_1 := \frac{1}{2\pi} \int_{\T^2_*} \mathcal{F} \quad \in \Z.
\end{equation}
This integer, called the \emph{Chern number}, is the topological invariant which underlies the quantization of the (anomalous) Hall conductivity in Chern insulators \cite{Haldane88,Experiment,Bestwick et al 2015,Chang et al 2015} and quantum Hall insulators \cite{TKNN,Graf review}. In the specific case under investigation of the Haldane Hamiltonian, the four regions of parameters $(\phi,M)$ in which $H(\ve{k})$ is gapped can be labelled by the Chern number \cite{Haldane88}: with reference to the colors of Figure \ref{PhaseDiagram}, the Chern number can be computed, \eg starting from \eqref{ChernSingularity}, to be $c_1=-1$ for the cyan region, $c_1=+1$ for the orange region, and $c_1=0$ for the two white regions. In analogy with the thermodynamical phases of statistical mechanics, one then speaks of \emph{topological phases of matter} distinguished by different topological invariants, and refers to Figure \ref{PhaseDiagram} as the \emph{topological phase diagram} for the Haldane Hamiltonian.

\begin{remark}
It is interesting to notice that the topological content associated to singularities of the Bloch function at the Dirac points persists, in an appropriate sense, also in the \emph{gapless} regime. If for example the parameters $(\phi,M)$ are threaded from the cyan region to the lower white region of Figure \ref{PhaseDiagram}, passing through a point in parameter space $(\phi_*,M_*)$ where $R_3(\ve{K})=0$, then at $(\phi_*,M_*)$ not only the Bloch function $u_-$ but also the projection $P_-$ becomes singular at $\ve{K}$. Nonetheless, the topological charge exchanged through the gapless phase can be quantified by means of a local topological invariant, the \emph{eigenspace vorticity}, associated to family of projections $P_-$ around the singular point $\ve{K}$ \cite{MonacoPanati2014}. In the situation described above, this eigenspace vorticity equals $\Delta c_1 = 0 - (-1) = 1$.
\hfill $\diamond$
\end{remark}

\goodbreak

\section{The localization dichotomy for periodic insulators}
\label{Sec:Dichotomy}

It is astounding to discover the predictive power of the Haldane model. In fact, it turns out that the features discussed in the previous Section are completely generic in two and three dimensions: indeed, the close connection between the structure of the singularities of the Bloch functions and the topology of the associated eigenspaces persists in a much wider context and for more general models. 

This was recently proved and quantified in a precise way in \cite{MonacoPanatiPisanteTeufel2018}. To formulate the main result, we need to set up the more general framework. Let $d\le 3$. The configuration space of a crystalline system is modeled by the space $X$, which can be either $\R^d$ or a $d$-dimensional crystalline structure (\eg the honeycomb structure presented in Section \ref{Sec:Haldane}): $X$ carries an action of the lattice $\Gamma \simeq \Z^d$ by translations, which is assumed to lift to translation operators $T_\gamma \in \mathrm{U}(L^2(X))$, $\gamma \in \Gamma$. 

Associated to these translation operators, there is a \emph{Bloch--Floquet--Zak transform} $\mathcal{U} \colon L^2(X) \to L^2_\tau(\B;L^2\sub{per}(Y)) \simeq \int_{\B}^{\oplus} L^2\sub{per}(Y) \di \ve{k}$, defined by
\begin{equation} \label{BF}
\left(\mathcal{U} \psi\right)(\ve{k},\ve{y}) := \sum_{\gamma \in \Gamma} \eu^{-\iu \ve{k} \cdot (\ve{y}-\gamma)} \, (T_\gamma \psi)(\ve{y}), \quad \ve{k} \in \B, \, \ve{y} \in Y,
\end{equation}
on suitable $\psi \in L^2(X)$. Here $\B$ stands for the fundamental cell of the dual lattice $\Gamma^*$ (the \emph{Brillouin zone} in the physics literature), $Y$ stands for the fundamental cell of the lattice $\Gamma$ (compare \eqref{cellY}), and $L^2_\tau(\B;L^2\sub{per}(Y))$ is the Hilbert space
\[ 
L^2_\tau(\B;L^2\sub{per}(Y)) := \left\{ 
\begin{gathered}
u \in L^2\sub{loc}(\R^d;L^2\sub{loc}(\R^d)) : \\
u(\ve{k}+\lambda,\ve{y}) = (\tau_\lambda u)(\ve{k},\ve{y}) := \eu^{-\iu \lambda \cdot \ve{y}} \, u(\ve{k},\ve{y}) \\
\text{and } T_\gamma u(\ve{k},\cdot) = u(\ve{k},\cdot) \\
\forall \ve{k} \in \R^d, \, \ve{y} \in \R^d, \, \lambda \in \Gamma^*, \, \gamma \in \Gamma 
\end{gathered} \right\} 
\]
of functions of the \emph{Bloch momentum} $\ve{k}$ and of the degrees of freedom in the unit cell $\ve{y}$ which are quasi-periodic (\emph{$\tau$-covariant}) in $\ve{k}$ and periodic in $\ve{y}$ (see \cite{MonacoPanati2015} for details). For crystalline structures of the type described in Section \ref{Sec:Haldane}, $\mathcal{U}$ coincides with the Fourier transform \eqref{Def:Fourier} up to the extra phase factor $\eu^{-\iu \ve{k} \cdot \ve{y}}$ in \eqref{BF}, which turns periodic functions of $\ve{k}$ into quasi-periodic, but makes the boundary conditions on the unit cell $Y$ in direct space $\ve{k}$-independent (namely, exactly periodic). A Bloch--Floquet--Zak transform is defined by \eqref{BF} also in the continuum case $X=\R^d$, where $T_\gamma$ can be the standard translation $(T_\gamma \psi)(\ve{y}) := \psi(\ve{y}-\gamma)$ or, more interestingly, a \emph{magnetic translation} generated by a uniform magnetic field with flux per unit cell which is commensurate to the flux quantum (equal to $2\pi$ in Hartree units), see \cite{Zak1964} and the discussion in \cite[Sec.~3]{MonacoPanatiPisanteTeufel2018}. Also in this case we will denote by $A \longleftrightarrow A(\ve{k})$ the correspondence between a periodic operator $A$ on $L^2(X)$ such that $[A,T_\gamma]=0$ for all $\gamma \in\ \Gamma$ and its decomposition into fibers in the Bloch--Floquet--Zak representation: $\mathcal{U} \, A \, \mathcal{U}^{-1} = \int_{\B}^{\oplus} A(\ve{k}) \, \di \ve{k}$.

Now that the framework of crystalline systems is clear, we can formulate the main hypothesis of the central result from \cite{MonacoPanatiPisanteTeufel2018}, which abstracts the predominant features of the Haldane Hamiltonian described in the previous Sections.

\medskip

\noindent \textbf{Assumption.} Let $H$ be a periodic self-adjoint operator on $L^2(X)$ with $H \longleftrightarrow H(\ve{k})$ where $H(\ve{k})$ defines a family of operators on $L^2\sub{per}(Y)$ such that
\begin{enumerate}
 \item $\set{H(\boldsymbol{\kappa})}_{\boldsymbol{\kappa} \in \C^d}$ defines an \emph{entire analytic family in the sense of Kato} with compact resolvent \cite{RS4};
 \item the family is \emph{$\tau$-covariant}, that is, $H(\ve{k}+\lambda) = \tau_{\lambda} \, H(\ve{k}) \, \tau_{\lambda}^{-1}$ for all $\ve{k} \in \R^d$ and $\lambda \in \Gamma^*$;
 \item the family is \emph{gapped}, namely there exists a set $\mathcal{I} \subset \N$ with $|\mathcal{I}| = m < \infty$ such that 
 \[ \inf_{\ve{k} \in \R^d} \inf_{\substack{n \in \mathcal{I} \\ m \in \N \setminus \mathcal{I}}} \big| E_n(\ve{k}) - E_m(\ve{k}) \big| \ge g > 0 \]
 where $\sigma(H(\ve{k}))=\set{E_n(\ve{k})}_{n \in \N}$ denotes the spectrum of $H(\ve{k})$ (consisting of discrete eigenvalues, the \emph{Bloch bands}, by the compact resolvent assumption).
\end{enumerate}

\medskip

In the discrete case (\eg for the Haldane Hamiltonian), the regularity assumption is easy to verify, as it is equivalent in position space to having sufficiently fast decaying hoppings between different sites of the crystal (say, exponential in the distance between the sites), and thus is in particular satisfied whenever the hopping Hamiltonian has finite range, as often happens in applications. For (magnetic, periodic) Schr\"{o}dinger operators, there are standard $L^p$-regularity assumptions on the electro-magnetic potentials that guarantee analyticity of the corresponding fiber Hamiltonians \cite{RS4}.

Notice moreover that the gap assumption allows to define the family of spectral projections $P(\ve{k})$ onto the spectral island $\sigma_0(\ve{k}) := \set{E_n(\ve{k}) : n \in \mathcal{I}}$, for example through the \emph{Riesz formula}
\[ P(\ve{k}) = \frac{\iu}{2 \pi} \oint_{C(\ve{k})} (H(\ve{k}) - z)^{-1} \di z, \]
where $C(\ve{k})$ is a positively oriented contour in the complex energy plane, locally constant in $\ve{k}$, which lies in the resolvent set of $H(\ve{k})$ and encircles only the eigenvalues in $\sigma_0(\ve{k})$. This family of projections is then $\tau$-covariant and depends analytically on $\boldsymbol{\kappa} \in \Omega_\alpha$, where $\Omega_\alpha \subset \C^d$ is a complex strip of half-width $\alpha>0$ around the ``real axis'' $\R^d \subset \C^d$ \cite[Prop.~2.1]{PanatiPisante}.

As in \eqref{Chern}, we can define the \emph{Chern numbers} associated to $\set{P(\ve{k})}_{\ve{k} \in \R^d}$ as
\begin{equation} \label{Chern_ij}
c_1(P)_{ij} := \frac{1}{2\pi} \int_{\B_{ij}} \Tr_{L^2\sub{per}(Y)} \left( P(\ve{k}) \left[ \partial_{k_i} P(\ve{k}), \partial_{k_j} P(\ve{k}) \right] \right) \di k_i \wedge \di k_j, \quad 1 \le i < j \le d,
\end{equation}
where $\B_{ij} \subset \B$ is the $2$-dimensional sub-torus of $\B$ where the coordinate different from the $i$-th and $j$-th is fixed (\eg to zero).

We are finally able to state the main result from \cite{MonacoPanatiPisanteTeufel2018}, generalizing the analysis on the Haldane Hamiltonian from the previous Section.

\begin{theorem}[{\cite{MonacoPanatiPisanteTeufel2018}}] \label{LocDic_Bloch}
Let $H \longleftrightarrow H(\ve{k})$ be as in the above Assumption, and $P(\ve{k})$ be the spectral projection onto the gapped spectral island of $H(\ve{k})$. Then for all $s \in [0,1)$ there exists a \emph{Bloch frame} $\set{u_1, \ldots, u_m} \subset H^s_\tau(\B;L^2\sub{per}(Y))$ for $\set{P(\ve{k})}_{\ve{k} \in \R^d}$, namely a set of functions $u_a \in H^s\sub{loc}(\R^d;L^2\sub{per}(Y))$ such that
\[ u_a(\ve{k}+\lambda) = \tau_\lambda u_a(\ve{k}), \quad \inner{u_a}{u_b}_{L^2} = \delta_{ab}, \quad \text{and} \quad P(\ve{k}) = \sum_{a=1}^{m} \ket{u_a(\ve{k})} \bra{u_a(\ve{k})}. \]

Moreover, the following statements are equivalent:
\begin{enumerate}
 \item there exists a Bloch frame in $H^1_\tau(\B;L^2\sub{per}(Y))$;
 \item there exists a Bloch frame in $C^\omega_\tau(\Omega_\alpha; L^2\sub{per}(Y))$, the space of $\tau$-covariant analytic functions on $\Omega_\alpha$ with values in $L^2\sub{per}(Y)$;
 \item the Chern numbers $c_1(P)_{ij}$, $1 \le i < j \le d$, defined in \eqref{Chern_ij}, vanish.
\end{enumerate}
\end{theorem}

The above result can be interpreted as a \emph{Localization--Topology Correspondence}, having implications also for the transport properties of the model under scrutiny for a crystalline insulator. To better clarify this point, we need to introduce one further notion. Given a periodic Hamiltonian $H \longleftrightarrow H(\ve{k})$ as in the Assumption above, denote by $P = \mathcal{U}^{-1} \left( \int_{\B}^{\oplus} P(\ve{k}) \di \ve{k} \right) \mathcal{U}$ the periodic projection on $L^2(X)$ onto the subspace corresponding to the isolated spectral island in momentum space. The Hamiltonian $H$ has generically absolutely continuous spectrum (which is given by $\sigma(H) = \left\{\lambda \in \R : \lambda = E_n(\ve{k}) \text{ for some } n \in \N, \, \ve{k} \in \R^d\right\}$), so it is not possible in general to find a basis of the range of $P$ given by eigenstates of the Hamiltonian. Nonetheless, if $\set{u_a(\ve{k})}_{1 \le a \le m}$ is an orthonormal basis for $\Ran P(\ve{k})$ --- a Bloch frame, in the terminology introduced above --- then it is possible to define (\emph{composite}) \emph{Wannier functions} \cite{MarzariEtAl12} by
\[ w_a(\ve{y}-\gamma) := (\mathcal{U}^{-1} u_a)(\ve{y}-\gamma) = \frac{1}{|\B|} \int_{\B} \eu^{\iu \ve{k} \cdot (\ve{y}-\gamma)} u_a(\ve{k},\ve{y}) \, \di \ve{k} , \quad 1 \le a \le m,  \]
where $\ve{y} \in Y$ and  $ \gamma \in \Gamma$.
The functions $w_a$ will automatically be in $\Ran P \subset L^2(X)$, and so will the translates $T_\gamma w_a$ by periodicity of $P$. One can then check \cite{Kuchment16} that $\set{T_\gamma w_a}_{\gamma \in \Gamma, \, 1 \le a \le m}$ constitutes an orthonormal basis for $\Ran P$ if the Bloch frame is $\tau$-covariant. \emph{Localized} Wannier functions are found to describe accurately the orbitals of the crystalline insulator \cite{MarzariEtAl12}, and it is hence important to understand their decay properties at infinity. Since the Bloch--Floquet--Zak transform shares with the standard Fourier transform the property of intertwining the multiplication operator by $\ve{x}$ on $L^2(X)$ and the gradient $\nabla_{\ve{k}}$ with respect to the crystal momentum, one can read off these decay properties of Wannier functions by looking at the smoothness with respect to $\ve{k}$ of the corresponding Bloch frame. More precisely, it holds that
\begin{align*}
&\langle \ve{x} \rangle^s w_a \in L^2(X) \quad &&\Longleftrightarrow \quad u_a \in H^s(\B;L^2\sub{per}(Y)), \quad s \ge 0, \\
&\eu^{\beta |\ve{x}|} w_a \in L^2(X) \; , \forall \beta \in [0,\alpha) &&\Longleftrightarrow \quad u_a \in C^\omega(\Omega_\alpha;L^2\sub{per}(Y)),  
\end{align*}
where we have denoted $\langle \ve{x} \rangle := (1 + |\ve{x}|^2)^{1/2}$.

The existence of a basis of well-localized (say, exponentially) Wannier functions signals the absence of charge transport in the crystal; on the contrary, a power-law decay of the Wannier functions is an indication of topological transport. If the Hall conductivity is non-zero, one then expects Wannier functions to be poorly localized. This is exactly the content of the above Theorem, which can be recast in terms of Wannier functions as a \emph{Localization Dichotomy}: \textsl{either} Wannier functions are exponentially localized (and this happens exactly when the Hall conductivity vanishes), \textsl{or} they are delocalized in the sense that they yield an infinite expectation of the squared position operator $|\ve{x}|^2$; no intermediate regimes of decay are allowed. The precise result is as follows.

\begin{theorem}[Localization Dichotomy {\cite{MonacoPanatiPisanteTeufel2018}}] \label{LocDic_WF}
Let $H$ be as in the above Assumption, and $P$ be the spectral projection onto the gapped spectral island. Then for all $s \in [0,1)$ there exists a \emph{Wannier basis} for $\Ran P$, that is, an orthonormal basis $\set{T_\gamma w_a}_{\gamma \in \Gamma, 1 \le a \le m}$ of $\Ran P$, such that
\[ \sup_{\gamma \in \Gamma} \int_{X} \langle \ve{x} - \gamma \rangle^{2s} \left|(T_\gamma w_a)(\ve{x}) \right|^2 \di \ve{x} \le C_s < \infty \quad \text{for all } s \in [0,1). 
\]

\smallskip

\noindent Moreover, the following statements are equivalent:
\begin{enumerate}
 \item there exists a Wannier basis such that
 \[ \sup_{\gamma \in \Gamma} \int_{X} \langle \ve{x} - \gamma \rangle^2 \left|(T_\gamma w_a)(\ve{x}) \right|^2 \di \ve{x} \le C_1 < \infty ; \]
 \item there exists a Wannier basis such that
 \[ \sup_{\gamma \in \Gamma} \int_{X} \eu^{2 \beta |\ve{x} - \gamma|} \left|(T_\gamma w_a)(\ve{x}) \right|^2 \di \ve{x} \le C_\omega < \infty \quad \text{for all } \beta \in [0,\alpha); \]
 \item the Chern numbers $c_1(P)_{ij}$, $1 \le i < j \le d$, defined in \eqref{Chern_ij}, vanish.
\end{enumerate}
\end{theorem}

We sketch here the main ideas from the proof of Theorem \ref{LocDic_Bloch}: for the detailed argument, the reader is referred to \cite{MonacoPanatiPisanteTeufel2018}. 

The first part consists in exhibiting a Bloch frame which is in $H^s_\tau$ for all $s \in [0,1)$. In $2d$, this is obtained via \emph{parallel transport}, a procedure which allows to construct a smooth ($C^\infty$) and $\tau$-covariant Bloch frame on the $1$-dimensional boundary of the Brillouin zone $\B$, and to extend this to the interior. The end result is a Bloch frame which is $\tau$-covariant and smooth except at one point in the Brillouin zone. The technique of parallel transport gives a precise control also on the type of singularity of the constructed Bloch frame, which is seen to be consistent with the claimed $H^s$-regularity (the derivatives of the Bloch functions have a $(1/r)$-divergence at the singular point). This situation should be compared with the Bloch function for the Haldane Hamiltonian exhibited in the previous Section. In $3d$, one needs to further extend an already singular datum at the $2$-dimensional boundary of the Brillouin zone to the $3$-dimensional ``bulk'': this can be done again by parallel transport, and produces this time lines of singularities, which dictate in turn the $H^s$-regularity in the statement of the Theorem.

The next part of the proof requires to show that if a Bloch frame in $H^1_\tau$ exists, then the Chern numbers of the family of projections vanish. The proof relies on a very subtle approximation of $H^1_\tau$ frames by $C^\infty_\tau$ frames. The subtlety lies in the fact that the space of frames in an Hilbert space is a non-linear manifold; the approximation of Sobolev maps with values in a manifold by regular maps becomes more involved, and requires in general certain topological conditions to be satisfied (see \cite{HangLin} and \cite[App.~B]{MonacoPanatiPisanteTeufel2018}). Nonetheless, in our setting $H^1$ maps can indeed be approximated by $C^\infty$ ones; when calculating an ``approximate'' Berry curvature \eqref{Berry} with the regular frames, its integrals over the tori $\B_{ij}$ are zero, so that in the limit the Chern numbers for the family of projections $P(\ve{k})$ must also vanish. It is then well-known \cite{Panati,PanatiPisante} how to modify the $H^1$-regular Bloch frame to an analytic one, provided the Chern numbers vanish.

As a side remark, note how in $2d$ the ``threshold'' Sobolev regularity $H^1$ coincides also with the ``threshold'' of the Sobolev embedding $H^s \hookrightarrow C^0$, which holds for $s>1$. Geometric arguments, based on the theory of vector bundles, yield that a non-zero Chern number forbids the existence of $\tau$-covariant \emph{continuous} Bloch frames \cite{Panati}: Theorem \ref{LocDic_Bloch} improves this result, claiming that also Bloch frames in $H^1_\tau$ cannot exist when the Chern numbers are non-vanishing. In $3d$, the result is even more stringent, as the threshold for the Sobolev embedding of $H^s$ into continuous functions is at $s=3/2$.

\section{The localization dichotomy for non-periodic insulators}
\label{Sec:GeneralizedDichotomy}
The results presented so far on the interplay between the localization properties and the topological features of crystalline systems exhibit a clear ``logical order'':  even though the Wannier functions are position-space objects, they are defined in terms of the Bloch functions, which are intrinsically $\mathbf{k}$-space objects. In the same way, the topological marker, namely the Chern number(s), is defined in terms of the $\mathbf{k}$-space fibering of the projection on the gapped part of the spectrum by means of the Bloch--Floquet--Zak transform. However, there is no physical reasons for the dichotomic behaviour illustrated by Theorem \ref{LocDic_WF} to hold only in systems for which a $\mathbf{k}$-space description is possible. Indeed, perfect crystalline systems do not exist in nature and an extension of the localization dichotomy to non-periodic gapped quantum systems is very tempting and desired. 

As a starting point for this ambitious goal, it is necessary to extend the notions of Wannier functions and Chern number to non-periodic systems. Distilling the true essence of Wannier functions, we get that they are a set of localized functions that, together with their copies obtained by lattice translation, form an orthonormal basis of a given spectral subspace, and encode the topological and transport information about the latter. Following this direction, we can extend the concept of Wannier basis to non-periodic systems.

In order to proceed further, we need the notion of \emph{localization function}. For the sake of the presentation, we restrict to $2$-dimensional continuum systems, namely we consider the Hilbert space $L^2(\R^2)$.
\begin{definition}[Localization function]
	\label{LocalizationFunction}
	We say that a continuous function $\Loc: [0,+\infty) \to (0,+\infty)$ is a \emph{localization function} if $\lim_{|\x| \to + \infty} \Loc (|\x|) \; = \; + \infty$ and there exists a constant $C_{\Loc}>0$ such that
	\begin{equation*}
	\label{GTriang}
	\Loc(|\x-\y|) \leq C_{\Loc} \; \Loc(|\x-\z|)  \; \Loc(|\z-\y|) \, , \qquad \forall \, \x,\y,\z \in \R^2 \, .
	\end{equation*}
\end{definition}

\noindent Following the seminal ideas in \cite{NenciuNenciu1993,NenciuNenciu1998}, we give a definition of generalized Wannier basis \cite{MarcelliMoscolariPanati,Moscolari}.
\begin{definition}[Generalized Wannier Basis]
	\label{GWB}
	Let $P \in \mathcal{B}(L^2(\R^2))$ be an orthogonal projection. Assume that there exist:
	\begin{enumerate}
		\item a Delone set $\Lattice \subseteq \R^2$, \ie a discrete set such that for some $0<r<R<\infty$ 
		it holds true that: 
		\begin{enumerate} 
			\item[(a)] for all $\x \in \R^2$ there is at most one element of $\Lattice$ in the ball of radius $r$ centered at $\x$ 
			(in particular, the set has no accumulation points);
			\item[(b)] for all $\x \in \R^2$ there is at least one element of $\Lattice$ in the ball of radius $R$ centered at $\x$ (the set is ``not sparse''); 
		\end{enumerate} 
		\item \label{L2norm} a localization function $\Loc$, constants $M >0$ and $m_* \in \N$ independent of $\gamma \in \Lattice$, and an orthonormal basis of $\Ran P$, denoted by $\{\w_{\gamma,a}\}_{\gamma \in \Lattice, 1 \leq a \leq m(\gamma)}$ with $m(\gamma) \leq m_* \; \forall \gamma \in \Lattice$, satisfying 
		\begin{equation*}
		\int_{\R^2} |\w_{\gamma,a}(\x)|^2 \, \Loc(|\x-\gamma|) \, \di \x \leq M  
		\end{equation*}
		for all $\gamma \in \Lattice, a  \in \set{1,\ldots, m(\gamma)}$. 
	\end{enumerate}
Then we call $\w_{\gamma,a}$ a \emph{generalized Wannier function} (GWF) with \emph{centre} $\gamma$, and we say that $P$ admits a \emph{generalized Wannier basis} (GWB) $\{\w_{\gamma,a}\}_{\gamma \in \Lattice, 1 \leq a \leq m(\gamma)}$. 
\end{definition}
When the localization function $\Loc$ is an exponential function, that is $\Loc(|\x|)=e^{2 \beta |\x|}$ for some $\beta > 0$,  
we say that the GWB is \emph{exponentially localized}. If the localization function $\Loc$ is of polynomial type, that is, $\Loc(|\x|)= \langle \ve{x}\rangle^{2s} = \big( 1+ |\x|^2 \big)^{s}$ for some $s > 0$,  we say that the GWB is \emph{$s$-localized}.

Notice that an orthonormal basis made of composite Wannier functions, as defined in Section \ref{Sec:Dichotomy},
is an example of GWB. Moreover, from the results in \cite{CorneanNenciuNenciu2008}, one concludes that for generic gapped $1$-dimensional systems there always exists an orthonormal basis for the range of the gapped spectral projection satisfying the requirements of Definition \ref{GWB} with the exception of some properties of the set $\Lattice$ (in particular $\Lattice$ is only proven to be a discrete set).  

As already mentioned, the Chern number has been historically related to the quantized conductivity in the QHE \cite{Graf review}. Since the experiments revealing the QHE have been realized with real materials, a generalization of the Chern number in presence of impurities and disorder has been considered long ago. Indeed, Bellissard, van Elst and Schulz-Baldes (see \cite{BellissardVanElstSchulzBaldes1994} and references therein), inspired by ideas in Non Commutative Geometry, extended the concept of Chern number to ergodic systems and connected it to the transverse conductivity in the QHE. Moreover, more recently, in the physics community there has been a growing interest in the analysis of topological materials by means of topological marker defined directly in position space \cite{BiancoResta2011, Caio et al 2019, Irsigler et al 2019}. Inspired by these ideas, we give the following definition of \emph{Chern marker} (which is called Chern character in \cite{CorneanMonacoMoscolari2018}, where the setting is slightly different).
\begin{definition}[Chern marker]
	\label{def:Cherncharacter}
	Let $P$ be a projection on $L^2(\R^2)$ and $\chi_{L}$ be the indicator function of the set $(-L,L]^2$. The \emph{Chern marker} of $P$ is defined by 
	\begin{equation*}
	C(P):=\lim\limits_{L \to \infty} \frac{2\pi}{4L^2} \Tr \Big(\iu\chi_{L} P \Big[\left[X_1, P \right],\left[X_2,P \right]\Big]P \chi_L\Big) 
	\end{equation*}
	
\noindent whenever the limit on the right hand side exists.
\end{definition}
Notice that, in case $P$ is an integral operator then the integral kernel of the operator $P [\left[X_1, P \right],\left[X_2,P \right]]P$ coincides with the definition of local Chern number given in \cite{BiancoResta2011} and with the definition of local Chern marker given in \cite{Caio et al 2019}.

Whenever the projection is periodic and hence can be fibered by the Bloch--Floquet--Zak transform, the Chern marker coincides with the Chern number, appearing in \eqref{Chern}, and hence it is an integer. Furthermore, one can show that the Chern marker is stable against regular perturbations and against the addition of a constant magnetic field, provided that these perturbations do not close the gap \cite{CorneanMonacoMoscolari2018}.

Hence, guided by the results in the periodic case, in \cite{MarcelliMoscolariPanati} the authors conjecture the existence of a relation between the localization properties of a GWB for a projection and the Chern marker of the projection itself. To formulate this conjecture properly, we focus on physical systems that can be described by a Hamiltonian operator $H$, acting in the Hilbert space $L^2(\R^2)$, and of the form
\begin{equation}
\label{Hamiltonian}
H=-\frac{1}{2}\Delta_{\bf A} + V \, ,
\end{equation}
where $V$ is a scalar potential such that $V$ is in $L^2\sub{u-loc}(\R^2)$, and $-\Delta_{\bf A}:=\left(-\iu \nabla-\mathbf{A}\right)^2$ is the magnetic Laplacian. The magnetic potential $\mathbf{A}$ is such that $\mathbf{A} \in L^4\sub{loc}(\R^2,\R^2)$ and the distributional derivative $\nabla \cdot \mathbf{A}$ is in $L^2\sub{loc}(\R^2)$. The assumptions on the potentials are the usual assumptions which allow to apply the diamagnetic inequality. Under these hypotheses, the Hamiltonian is essentially selfadjoint on the dense core $C^\infty_0(\R^2)$. Moreover, we assume that the spectrum of the Hamiltonian has a spectral island $\sigma_0(H)$ isolated from the rest of the spectrum of $H$, that is ,
$$
\dist(\sigma_0(H),\sigma(H)\setminus \sigma_0(H))= g > 0 \, .
$$

\noindent We can now formulate the

\medskip

\noindent \textbf{Localization Dichotomy Conjecture.}
{\it 
	Under the above assumptions, let $P$ be the spectral projection onto the spectral island $\sigma_0(H)$ of a Hamiltonian operator of the form \eqref{Hamiltonian}. Then the following statements are equivalent:
	\begin{enumerate}[label=(\alph*),ref=(\alph*)]
		\item \label{exp-loc} $P$ admits a generalized Wannier basis that is exponentially localized.
		\item \label{s-loc} $P$ admits a generalized Wannier basis that is $s_*$-localized for $s_*=1$.
		\item \label{c-zero} $P$ is topologically trivial in the sense that its Chern marker $C(P)$ exists and is equal to zero.
	\end{enumerate}
}

\medskip

Notice that \ref{exp-loc} easily implies \ref{s-loc} by a simple inequality, while the opposite implication is not trivial. The Localization Dichotomy Conjecture generalizes the results proved in the periodic setting in Theorem \ref{LocDic_WF} \cite{MonacoPanatiPisanteTeufel2018}. 

A new result, still unpublished, covering the non-periodic setting is the following \cite{Moscolari,MarcelliMoscolariPanati}. 
While we expect the Conjecture to be true for $s_*=1$, the theorem is restricted to $s_*>5$ for technical reasons.
\begin{theorem}[Localization implies topological triviality]
	\label{LocDicTheorem}
	Under the above assumptions, let $P$ be the spectral projection onto the spectral island $\sigma_0(H)$ of a Hamiltonian operator of the form \eqref{Hamiltonian}. Suppose that $P$ admits a $s_*$-localized generalized Wannier basis $\{\w_{\gamma,a}\}_{\gamma \in \Lattice, 1 \leq a \leq m(\gamma)}$ 
	for $s_*>5$, that is: there exists $M > 0$ and $m_* \in \N$ such that  
	\begin{equation*}
	\int_{\R^2} |\w_{\gamma,a}(\x)|^2 (1+\|\x-\gamma\|^2)^{s_*} \, d\x \leq M\, , \qquad \forall\; \gamma \in \Lattice \, , \forall a \in \left\{1, \dots, m(\gamma)\right\} 
	\end{equation*}
	with $m(\gamma)\leq m_*$ for all $\gamma \in \Lattice$.
	Then the Chern marker of $P$ is zero, namely the following limit exists and 
	\begin{equation*} \label{ZeroCN}
	\lim\limits_{L \to \infty} \frac{2\pi}{4L^2} \Tr \left(\iu\chi_{L} P \left[\left[X_1, P \right],\left[X_2,P \right]\right]P \chi_L\right)  \;=\; 0.
	\end{equation*}
\end{theorem}

The proof of Theorem \ref{LocDicTheorem} is obtained in two steps: first one proves exponential localization estimates for the integral kernel of the projection and for some auxiliary operators by using Combes-Thomas-type estimates. Then, by using those estimates, one can gain an explicit control on the asymptotic behaviour of the trace of $ \iu\chi_{L} P \left[\left[X_1, P \right],\left[X_2,P \right]\right]P \chi_L$ as $L \to \infty$ and therefore can prove the desired limit. Notice that the threshold $s_*>5$ is only due to technical reasons and, as we mentioned before, a full generalization of the localization dichotomy proved in \cite{MonacoPanatiPisanteTeufel2018} would require $s_*=1$, as well as a proof of the opposite implication. Further investigations are planned for the future in order to move the threshold to $s_*=1$.  

\begin{remark}
As a by-product of Theorem \ref{LocDicTheorem}, it follows that the dichotomic behaviour of the Wannier basis in Theorem \ref{LocDic_WF} is ``stable'' with respect to regular perturbations. Indeed, consider a periodic system such that its Chern number is different from zero and suppose that we perturb the system with a small non-periodic term, for example by adding some impurities modelled by Coulomb potentials. By contradiction, suppose that the perturbed system has an exponentially localized GWB in the sense of Definition \ref{GWB}. By a result proved by A. Nenciu and G. Nenciu \cite{NenciuNenciu1993}, it is possible to unitarily transport the GWB back to the original system. Then, Theorem \ref{LocDicTheorem} implies that the Chern marker is zero. As we have mentioned before, for periodic systems the Chern marker equals the Chern number. Therefore, this implies that the original periodic system has a vanishing Chern number and yields a contradiction.
\hfill $\diamond$
\end{remark}

Despite a proof of the implication \ref{c-zero} $\Rightarrow$ \ref{exp-loc} is still missing in the non-periodic setting, Theorem \ref{LocDicTheorem} provides a clear relation between the GWB and the Chern marker. Whenever a sufficiently localized GWB for a given gapped quantum system exists, one can be sure that such physical system does not exhibit Hall transport. This relation is completely independent of the periodicity of the system.



\bigskip \bigskip


\begin{thebibliography}{OO}



\bibitem{Ando} 
Ando, Y.: Topological insulator materials. J. Phys. Soc. Jpn. {\bf 82}, 102001 (2013).

\bibitem{BiancoResta2011}
Bianco, R., Resta, R.: Mapping topological order in coordinate space. Phys. Rev. B {\bf 84}, 241106 (2011).

\bibitem{BellissardVanElstSchulzBaldes1994} 
Bellissard, J., van Elst, A., Schulz--Baldes, H.:  The noncommutative geometry of the quantum Hall effect. 
J. Math. Phys. {\bf 35}, 5373--5451 (1994). 

\bibitem{BenaMontambaux} 
Bena, C., Montambaux, G.: Remarks on the tight-binding model of graphene. New J. Phys. {\bf 11}, 095003 (2009).

\bibitem{Bestwick et al 2015}
Bestwick, A.\,J., Fox, E.J., Kou, X., Pan,L., Wang, K.\,L., Goldhaber-Gordon, D.: Precise quantization of the anomalous  Hall effect near zero magnetic field. Phys. Rev. Lett. {\bf 114}, 187201 (2015). 

\bibitem{Caio et al 2019}
Caio, M.\,D., M\"oller, G., Cooper, N.\,R., Bhaseen, M.\,J.: Topological marker currents in Chern insulators. 	Nat. Phys. {\bf 15}, 257--261 (2019). 

\bibitem{CorneanMonacoMoscolari2018}
Cornean, H., Monaco, D., Moscolari, M.: Beyond Diophantine Wannier diagrams: gap labelling for Bloch-Landau Hamiltonians. \href{https://arxiv.org/abs/1810.05623}{\texttt{arXiv:1810.05623 [math-ph]}} (2018). 

\bibitem{CorneanNenciuNenciu2008} 
Cornean, H.D., Nenciu, A., Nenciu. G.: Optimally localized Wannier functions for quasi one-dimensional nonperiodic insulators. J. Phys. A Math. Theor. {\bf 41}, 125202 (2008).

\bibitem{Experiment}
Chang, C.\,Z. \textsl{et al.}: Experimental Observation of the Quantum Anomalous Hall Effect in a Magnetic Topological Insulator. Science {\bf 340}, 167--170 (2013).

\bibitem{Chang et al 2015}
Chang, C.\,Z. \textsl{et al.}: High-precision realization of robust quantum anomalous Hall state in a hard ferromagnetic topological insulator. Nat. Mat. {\bf 14}, 473 (2015).

\bibitem{FruchartCarpentier2013}
Fruchart, M., Carpentier, D.: An introduction to topological insulators. C.\,R.\,Phys. {\bf 14}, 779--815 (2013).

\bibitem{Fruchart2014}
Fruchart, M., Carpentier, D., Gawedzki, K.: Parallel Transport and Band Theory in Crystals. EPL {\bf 106}, 60002 (2014).

\bibitem{GiulianiMastropietroPorta2017}
Giuliani, A., Mastropietro, V., Porta, M.: Universality of Hall conductivity in interacting electron systems. Commun. Math. Phys. {\bf 349}, 1107--1161 (2017).

\bibitem{Graf review}
Graf, G.\,M.: Aspects of the Integer Quantum Hall Effect. In: Spectral Theory and Mathematical Physics: A Festschrift in Honor of Barry Simon's 60th Birthday, No. 76 in Proceedings of Symposia in Pure Mathematics, pages~429--442. American Mathematical Society, Providence (2007).

\bibitem{Haldane88}
Haldane, F.\,D.\,M.: Model for a Quantum Hall Effect without Landau levels: condensed-matter realization of the ``parity anomaly''. Phys. Rev. Lett. {\bf 61}, 2017 (1988).

\bibitem{HangLin} 
Hang, F., Lin, F.\,H.: Topology of Sobolev mappings II. Acta Math. {\bf 191}, 55--107 (2003).

\bibitem{HasanKane}
Hasan, M.\,Z., Kane, C.\,L.: Colloquium: Topological Insulators. Rev. Mod. Phys. {\bf 82}, 3045--3067 (2010).

\bibitem{Hofstadter76} 
Hofstadter, D.\,R.: Energy levels and wave functions of Bloch electrons in rational and irrational magnetic fields. Phys. Rev. B {\bf 14}, 2239--2249 (1976).

\bibitem{Irsigler et al 2019}
Irsigler, B., Zheng, J., Hofstetter, W.: Microscopic characteristics and tomography scheme of the local Chern marker. Phys. Rev. A {\bf 100}, 23610 (2019).


\bibitem{Kuchment16}
Kuchment, P.: An overview of periodic ellipic operators. Bull. AMS {\bf 53}, 343--414 (2016).

\bibitem{MarcelliMoscolariPanati}
Marcelli, G., Moscolari, M., Panati, G.: Localization of a generalized Wannier basis implies Chern triviality in non-periodic insulators. In preparation (2019).

\bibitem{MarzariEtAl12}
Marzari, N., Mostofi, A., Yates, J., Souza, I., Vanderbilt, D.: Maximally localized Wannier functions: Theory and applications. Rev. Mod. Phys. {\bf 84}, 1419--1475 (2012).

\bibitem{MonacoPanati2014}
Monaco, D., Panati, G.: Topological invariants of eigenvalue intersections and decrease of Wannier functions in graphene. J. Stat. Phys. {\bf 155}, 1027--1071 (2014).

\bibitem{MonacoPanati2015}
Monaco, D., Panati, G.: Symmetry and localization in periodic crystals: triviality of Bloch bundles with a fermionic time-reversal symmetry. Acta Appl. Math. {\bf 137}, 185--203 (2015).  

\bibitem{MonacoPanatiPisanteTeufel2018}
Monaco, D., Panati, G., Pisante, A., Teufel, S.: Optimal decay of Wannier functions in Chern and Quantum Hall insulators. Comm. Math. Phys. {\bf 359}, 61--100 (2018).

\bibitem{Moscolari}
Moscolari, M.: On the localization dichotomy for gapped quantum systems. PhD Thesis, Sapienza - University of Rome (2018).

\bibitem{NenciuNenciu1993} 
Nenciu, A., Nenciu. G.: The Existence of Generalised Wannier Functions for One-Dimensional Systems. Phys. Rev. B {\bf 47}, 10112--10115  (1993). 

\bibitem{NenciuNenciu1998} 
Nenciu, A., Nenciu. G.: Existence of exponentially localized Wannier functions for nonperiodic systems. Commun. Math. Phys. {\bf 190}, 541--548 (1998).

\bibitem{Panati}
Panati, G.: Triviality of Bloch and Bloch-Dirac bundles. Ann. Henri Poincar\'e {\bf 8}, 995--1011 (2007).

\bibitem{PanatiPisante}
Panati, G., Pisante, A.: Bloch bundles, Marzari-Vanderbilt functional and maximally localized Wannier functions. Commun. Math. Phys. {\bf 322}, 835--875 (2013).

\bibitem{RS4}
Reed M., Simon, B.: Methods of Modern Mathematical Physics. Volume IV: Analysis of Operators. Academic Press, New York (1978).

\bibitem{Santoro_LN} 
Santoro, G.\,E.: Lectures notes on Non-equilibrium quantum systems. \url{http://indico.ictp.it/event/7644/material/2/2.pdf} (accessed December 10, 2018). 


\bibitem{ThonhauserVanderbilt}Thonhauser, T., Vanderbilt, D.: Insulator/Chern-insulator transition in the Haldane model. Phys. Rev. B. {\bf 74}, 235111 (2006).

\bibitem{TKNN}
Thouless, D.\,J., Kohmoto, M., Nightingale, M.\,P., de Nijs, M.: Quantized Hall conductance in a two-dimensional periodic potential. Phys. Rev. Lett. {\bf 49}, 405--408 (1982).

\bibitem{Zak1964}
Zak, J.:  Magnetic translation group. Phys. Rev. {\bf 134}, A1602 (1964). 

\end{thebibliography}
\end{document}